\documentclass{aa}  

\usepackage{graphicx}
\usepackage{txfonts}
\usepackage{subcaption}         
\usepackage{lscape}             
\usepackage{placeins}           
                                
\usepackage{xspace}
\usepackage{tablefootnote}
\usepackage{threeparttable}
\usepackage{siunitx} 
\usepackage{booktabs} 
\usepackage{hhline} 
\usepackage{amsmath} 
\usepackage{hyperref}
\usepackage{caption}

\numberwithin{equation}{section}

\newcommand{\cygx}{Cygnus X-1\xspace}

\newcommand{\Ecut}{2.8}
\newcommand{\Eplus}{0.6}
\newcommand{\Eminus}{0.8}
\newcommand{\kte}{44.7}
\newcommand{\kteplus}{1.9}
\newcommand{\kteminus}{1.7}
\newcommand{\taup}{1.10}
\newcommand{\tauperr}{0.06}
\newcommand{\elecgev}{20.6}
\newcommand{\elecgevplus}{2.8}
\newcommand{\elecgevminus}{2.2}
\newcommand{\ulow}{0.24}
\newcommand{\uhigh}{0.75}
\newcommand{\gammashlow}{1.1}
\newcommand{\gammashhigh}{2.0}

\newcommand{\helicity}{13}

\begin{document}

    \title{A unified modeling of X-ray and gamma-ray spectro-polarimetric data: the case of Cygnus X-1}
    

   \author{T. Bouchet\inst{\ref{inst:jmu}} \and
            T. Siegert\inst{\ref{inst:jmu}} \and
            V. Grinberg\inst{\ref{inst:esa}} \and
            J. Rodriguez\inst{\ref{inst:cea},\ref{inst:osups}} \and
            F. Cangemi\inst{\ref{inst:apc}} \and
            P. Laurent\inst{\ref{inst:cea}} \and
            J. Wilms\inst{\ref{inst:fau}} 
          }

  \institute{
  {Julius-Maximilians-Universität Würzburg, Fakultät für Physik und Astronomie, Institut für Theoretische Physik und Astrophysik, Lehrstuhl für Astronomie, Emil-Fischer-Str. 31, 97074 Würzburg, Germany\label{inst:jmu}}
  \and
  {European Space Agency (ESA), European Space Research and Technology Centre (ESTEC), Keplerlaan 1, 2201 AZ Noordwijk, The Netherlands\label{inst:esa}}
  \and
  {Universit\'e Paris-Saclay, Universit\'e Paris Cit\'e, CEA, CNRS, AIM, F-91191 Gif-sur-Yvette, France\label{inst:cea}}
  \and
  {Observatoire des Sciences de l’Univers de l’université Paris-Saclay, Bat 121, F-91405 Orsay, France\label{inst:osups}}
  \and
  {APC, Universit\'e Paris Cit\'e/CNRS/CEA, 75013 Paris, France \label{inst:apc}}
  \and
  {Dr.\ Karl Remeis-Observatory, Friedrich-Alexander-Universit\"at Erlangen-N\"urnberg, Sternwartstr.~7, 96049 Bamberg, Germany \label{inst:fau}}
}

   \date{received, accepted}

 
  \abstract
   {With the advent of new instrumentation, the available spectro-polarimetric measurements in the X-ray and gamma-ray domains have increased significantly. For black hole X-ray binaries, such as Cygnus X-1, a clear discrepancy between the independent measurements of X-rays and soft $\gamma$-rays polarization has emerged.}
  {This discrepancy challenges our understanding of the accretion process and emission mechanisms, including acceleration processes at play and the geometry of the system. We thus develop a general purpose method to model the broadband high-energy spectro-polarimetric data.}
   {Our novel approach allows to simultaneously fit the spectral and polarization data from independent instruments in different energy bands. Measurements with widely different energy bin sizes and statistical significances are combined in a robust and statistically sound way in a single fit, taking into account energy-dependent polarization mixing from different model components.}
   {We apply our method to the spectro-polarimetric data of the black hole X-ray binary Cygnus X-1 in the 2\,keV--2\,MeV range and show that it enables us to describe the shape of the elusive soft $\gamma$-rays hard-tail with unprecedented quality and details. We constrain the photon cut-off energy from optically thin synchrotron emission at $(\Ecut^{+\Eplus}_{-\Eminus}) \times 10^{2}\ \mathrm{keV}$. From this value, we discuss the origin of the non-thermal electrons in the context of Bohm diffusion and synchrotron cooling. We further provide a new explanation for the misalignment of the polarization of the synchrotron hard-tail compared to the jet axis, based on magnetic field helicity and Doppler boosting. Overall, our new approach offers to the broader community a simple and straightforward way to apply models to diverse spectro-polarimetric data.}
   {}


   \maketitle

\section{Introduction}

Recently, polarization in the X-ray and $\gamma$-ray bands have received much attention, due to an increase in the number of independent measurements \citep{Chattopadhyay_2021, Trippe_2014, ixpe_main_results}. The interpretation of such measurements can bring new constraints on the physics of astrophysical sources, especially when combined together \citep{Russell_2014, Liodakis_2022, Ursini_2023, 2023MNRAS.519.5902G, Farinelli_2025}. In many cases, the spectral and polarized information for a single source can potentially come from different instruments, with widely different data analyses associated. 
Although many advanced modeling approaches have been developed \citep[e.g.][]{Kislat_2015, Strohmayer_2017}, they are not modulable enough as they work best for data of similar statistical weights.
This issue is aggravated since the energy binning is much coarser and statistical uncertainties are much larger for polarization than for spectroscopic data and for $\gamma$-rays than for X-rays data.
Here, we present a novel approach that combines in a straightforward way the spectro-polarization analysis in most of the high-energy domain, where the polarization here is the simultaneous measurement of the Polarization Angle (PA) and Polarization Fraction (PF). Unlike other methods, this allows us to include the soft $\gamma$-ray spectro-polarization ($E\gtrsim100$\,keV) information, which is often left out despite its impact on the global understanding of the astrophysical processes \citep{Zdziarski_2014}.

\subsection{BHXB polarization}
\begin{figure}
    \centering
    \includegraphics[width=\linewidth]{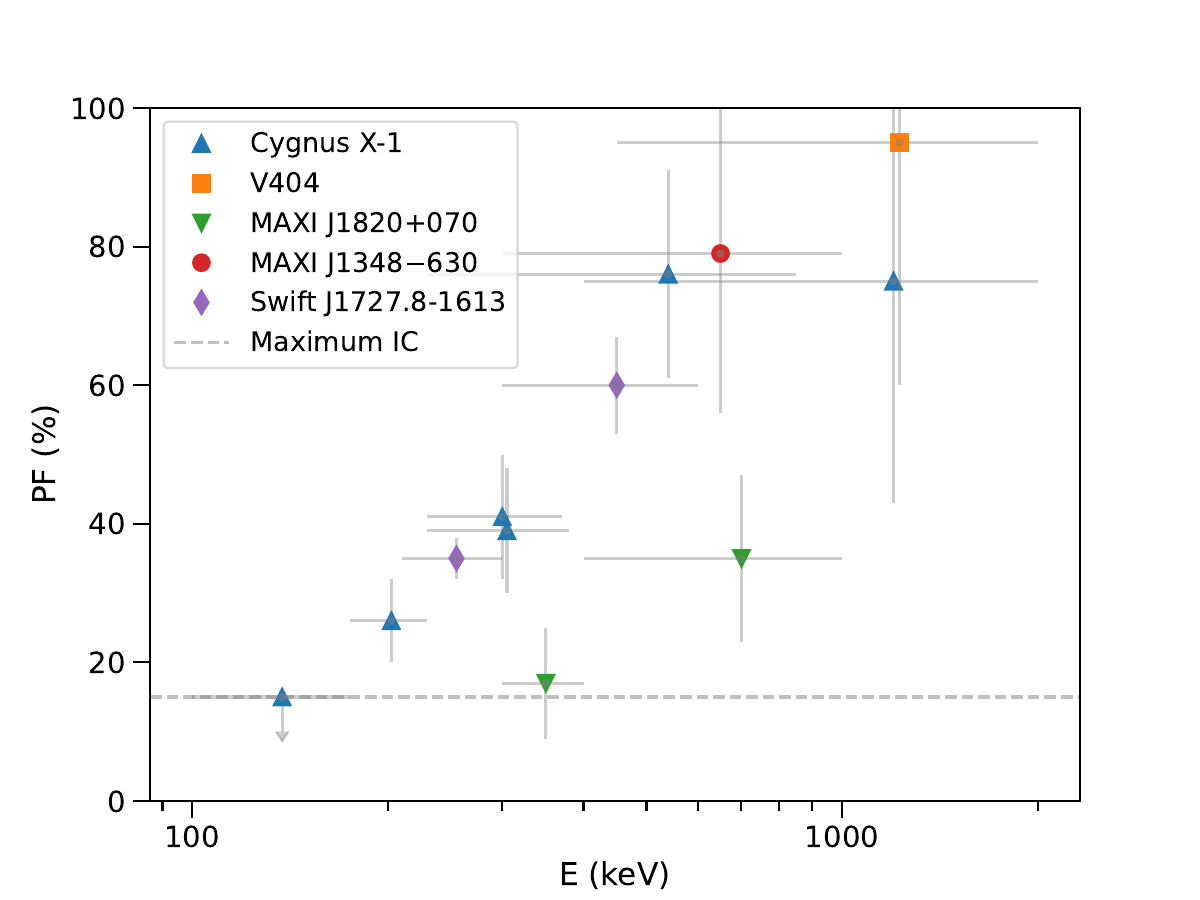}
    \caption{All previous measurements of the PF of BHXB hard-tails in the hard states, above 100\,keV. The maximally allowed polarization for IC is shown as a grey dotted line.}
    \label{fig:historic_mq_pf}
\end{figure}

Black hole X-ray binary (BHXB) are among the brightest soft $\gamma$-ray sources, making them the perfect candidates to showcase our method \citep{bhxb_integral_review}.
In BHXBs, the thermal emission from an accretion disk is believed to explain the emission near 1\,keV. An additional component dominates up to $\sim$100\,keV, likely coming from the inverse Comptonization (IC) of low energy seed photons by a thermal population of electrons \citep{Sunyaev_1980}. BHXBs also experience state transitions, between a (low) hard state (HS), dominated by the IC, and (high) soft state, dominated by the thermal emission \citep{Gierlinski_1997, Belloni_2010}. Above the thermal cut-off of the electrons, a high-energy tail has been observed in most of the brightest BHXBs, such as \cygx \citep{McConnell_2000, Laurent_2011, Jourdain_2012_polar}, 1E\,1740.7$-$2942 \citep{Bouchet_1991},  GX\,339$-$4 \citep{gx339_ht}, GRS 1915+105 \citep{Rodriguez_2008}, V404\,Cyg \citep{Rodriguez_2015v404, Siegert_2016, Jourdain_2017}, Cygnus X-3 \citep{cygx3_ht}, MAXI\, J1820+070 \citep{maxij1820_ht}, MAXI\, J1535$-$571, MAXI\, J1348$-$630 \citep{Cangemi_2023a}, and Swift\, J1727.8$-$1613 \citep{Bouchet_2024}. This hard $\gamma$-tail is explained by a population of non-thermal electrons from which the emission process could be IC \citep{Zdziarski_1993, Poutanen_hybrid, Malzac_2009}, synchrotron \citep{Laurent_2011, Zdziarski_2012}, or pair plasma annihilation \citep{1999MNRAS.305..181B, 1999ApJ...510L.123B}. Polarization measurements are particularly well-suited to distinguish between those scenarios. So far, polarization detections strictly above 100\,keV have only been possible in the hard states of the brightest BHXB (see Fig. \ref{fig:historic_mq_pf}).
In those cases, synchrotron emission from a highly ordered magnetic field has been the preferred explanation for most of the high PF (PF$\gtrsim20\%$) observed \citep{Rodriguez_2015, Laurent_2016, Cangemi_2023a, Bouchet_2024, Chattopadhyay_2024} since in the most realistic configurations, IC indeed only reaches limited PF below 15\% for typical BHXB geometries \citep{hybrid_corona_polar}.\\

\cygx is a high-mass X-ray binary (HMXB) where the BH is fed from the dense stellar wind of its supergiant companion, in addition to a small accretion disk. For this reason, it is a persistent source that never goes into quiescence, spending most of its time in the HS, and transiting briefly in the soft state \citep{Grinberg_2015}.
In this source, the high degree of polarization was confirmed by three independent instruments in soft $\gamma$-rays: INTEGRAL/IBIS \citep[400--2000 keV range,][]{Laurent_2011, Rodriguez_2015}, INTEGRAL/SPI \citep[230--850 keV range,][]{Jourdain_2012_polar}, and more recently AstroSat/CZTI \citep[175--380 keV range,][]{Chattopadhyay_2024}. The lower energies were found to have much lower polarization (PF $< 10\%$), both in hard X-rays in the 19--181\,keV range \citep{Chauvin_2018, xlcalibur_2025} and soft X-rays in the 2--8\,keV range \citep{Krawczynski_2022_cygnus}.
Another surprising difference is the PA projected on the sky, which is aligned with the jet axis in X-rays \citep[$-$21°,][]{Krawczynski_2022_cygnus}, but has a wide shift in soft $\gamma$-rays \citep[42°,][]{Jourdain_2012_polar}.
Soft X-rays polarization was also shown to vary with states, with a lower PF in the soft state, most likely due to the accretion disk contribution \citep{Jana_2024, Steiner_2024,  Kravtsov_2025}.\\

\subsection{Polarization modeling}

The main scientific goal of this study is to understand this elusive hard-tail from spectro-polarization data, using the lower energies ($E\lesssim100$\,keV) as an additional constraint.
Within this framework, one important, and often overlooked aspect of polarization modeling, is the non-linearity arising from the superposition of various polarized sources, i.e. polarization mixing. The X-rays and $\gamma$-rays are particularly vulnerable to this problem, as these energy ranges combine different components, with a priori different fluxes, PAs and PFs. One important goal of this article is to explicitly show how the spectrum and polarization of different components can mix in non trivial way, how to take these effect into account rigorously, and how it applies to the concrete case of \cygx.\\

Few attempts have been made to study broadband polarization properties. Some qualitative predictions have been made from spectral-only modeling by \cite{Zdziarski_2012}, who showed that optically thin synchrotron emission could explain the MeV polarization for a hard power-law index ($p=1.4$) of the underlying electron distribution. Later, the same authors made a more thorough spectral modeling including GeV photons, and found that the high polarization is possible only for jet magnetic fields much above the equipartition level \citep{Zdziarski_2014}. \cite{Russell_2014} made the closest attempt to model both the spectrum and polarization from radio to soft $\gamma$-ray, although the PA seen in $\gamma$-rays was ignored by their model, and was only discussed later on its own. At the time when their paper was written, much fewer polarization measurements were available at X-ray and $\gamma$-ray energies. Especially in the 2--8\,keV band, polarization measurements only became available after the IXPE launch \citep{Weisskopf2022IXPE}.\\

Those previous studies can be improved by including all the relevant data when modeling and fitting for physical parameters, including variations of the PA and PF with energy. To proceed, two issues need to be addressed: Firstly, the spectrum and polarization are intertwined and influence each other in a non-trivial way so that a careful treatment of large energy bins and regions with overlapping emission components is necessary (Section \ref{sec:method}). Secondly, the general assumption that synchrotron polarization is constant with energy in a uniform magnetic field only holds for power-law electron distribution. For other distributions realistically expected for our sources \citep{cutoff_index_bohm, cutoff_shapes}, the polarization at high-energies is more complex (Section\,\ref{sect:models}). We apply our method to the case of \cygx using data from the literature (Section\,\ref{sec:data}). The model parameters are fitted with a least-square method first, and refined through a MCMC algorithm (Section\,\ref{sec:result}). Finally, the resulting parameters are interpreted and compared with previous studies (Section\,\ref{sec:interpretation}).

\section{Method}\label{sec:method}

In hard X-rays, polarization parameters are typically measured in broad energy bands \citep{Laurent_2011, Chauvin_2018, xlcalibur_2025}. In soft X-rays, the polarization per energy is converted into Stokes spectra and modeled \citep{Kislat_2015}. Applied to energies above 100\,keV, the model would either be too simple or not converge, since there are only a few data points available containing multiple components. Instead, the spectra that are defined on much finer binning have to be included along the polarization data in order to model correctly the spectral changes within broad continuum bins. Here, we detail the procedure we put in place to fit a model, taking into account the non-trivial way the flux and polarization from the different components mix. The main idea is to create a grid where the spectrum and polarization is computed as a function of energy, and then integrate this grid over the large polarization data bins. In the following, lower indices indicate a specific binning, and upper $m$ index indicates a predicted quantity.

\subsection{Spectrum}

As for all high-energy experiments, the raw spectrum are given in count-rates $C_s$, with an associated uncertainty $\sigma_s$ for each energy bin, $[E_s, E_s+\Delta E_s]$. Our model is composed of different components indexed by $k$ coming from different emission mechanisms. The predicted fluxes for each component ($F^m_{k}$) are calculated, summed, folded with the response matrix of the instrument, and re-binned within the 3ML Python library \citep{threeml_main}; this results in the predicted count-rates $C^m_s$ in the energy bins of the data. This is the classic "forward-modeling" approach, where the model-data comparison is done in the data space.

\subsection{Example of polarization mixing}\label{subsec:polar_mixing_example}

\begin{figure}
    \centering
    \includegraphics[width=\linewidth]{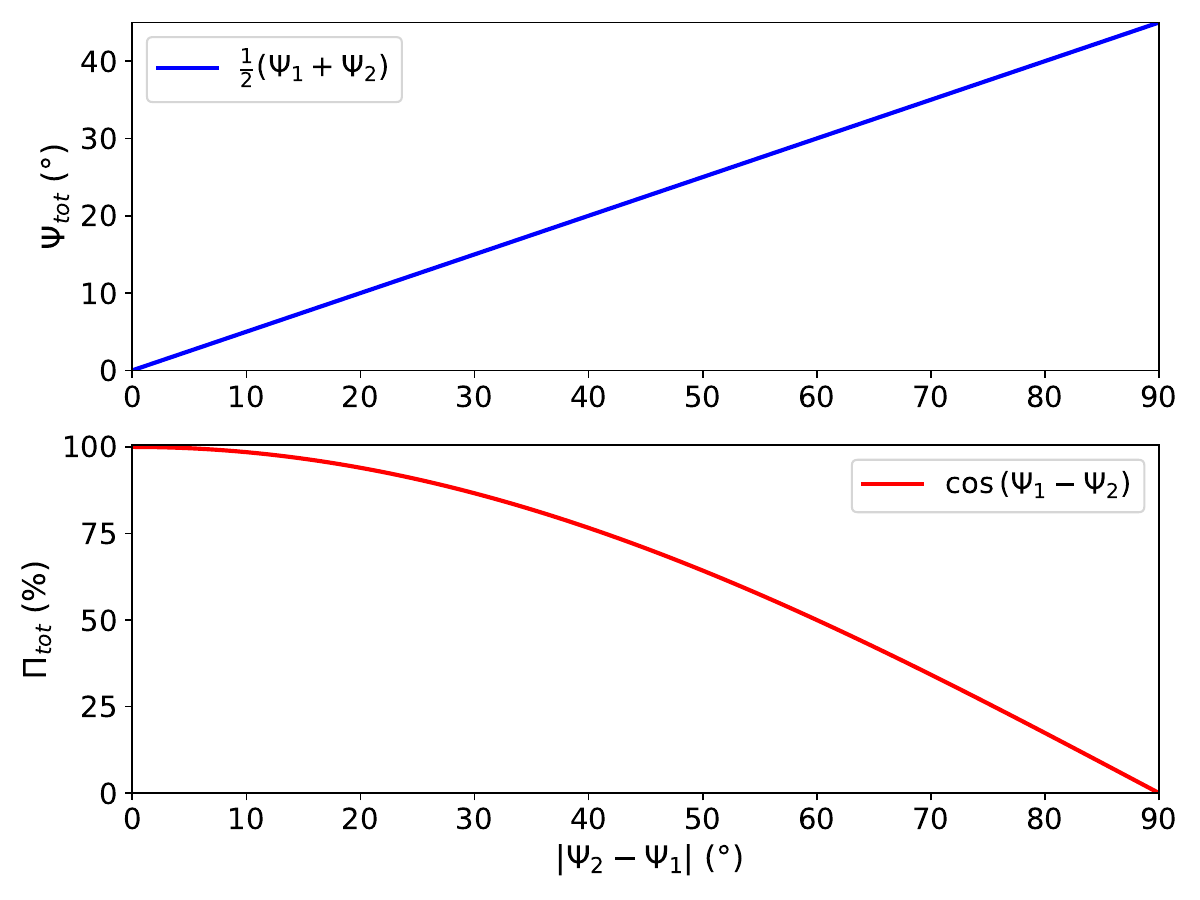}
    \caption{Summed emission of two polarized components, identical in every way except for their PA. The resulting PA (top) and PF (bottom) is on the y-axis, and the difference between the two component's PA on the x-axis.}
    \label{fig:pa_mixing_example}
\end{figure}

Before delving into the polarization mixing, we start with an intuitive example. Fig.\ref{fig:pa_mixing_example} shows the polarization of the combined emission from two identical components (1 and 2), with the same intensity ($I$), same PF ($\Pi=1$), but different PA ($\Psi_1$ and $\Psi_2$). As the difference in angle grows towards orthogonal, the resulting PF decreases towards 0\%, while the PA is always the average of the two angles being mixed.
One way to prove this is to use the complex representation of linear polarization, $P=I\;\Pi\,e^{2i\Psi}$, which is additive. The addition of the two complex polarizations gives,
\begin{equation}
    P_{tot} = Ie^{2i\Psi_1} + I e^{2i\Psi_2} = 2I \cos{(\Psi_1-\Psi_2)} e^{ i(\Psi_1+\Psi_2)}.
\end{equation}
Since $I_{tot} = 2I$, we can identify the PA and PF as,
\begin{equation}\label{eq:mixing_2_component}
    \left\{
    \begin{aligned}
    \Psi_{tot} &= \frac{\Psi_1+\Psi_2}{2} \\
    \Pi_{tot} &= \cos{(\Psi_1-\Psi_2)}
    \end{aligned}
    \right.
\end{equation}
This confirms the intuition that the resulting polarization of a mix of polarized emissions with different PA will add "destructively" if their difference is large, and always result in an average of both PAs. In reality, the flux (or intensity) also plays a weighting role; which component dominates the polarization is therefore roughly determined by the product $\Pi \times I$, as we will show in the following calculations.\\
 
\subsection{Polarization mixing algorithm}\label{subsec:polar_mixing}

In our method, the polarization model is not folded with the instrument response, since it is usually not given for each instrument, unlike for spectra. It is therefore required to convert the measurements into the Stokes parameters ($I,Q,U,V$) for all instruments, which are additive, and have approximately Gaussian uncertainty \citep{radiative_process_astro}. Circular polarization is not measured, and assumed to be negligible, therefore $V$ is 0 \citep[][]{polarization_astro_book}. The polarization data points, indexed by $j$, are taken from various instruments within energy bins $[E_j,E_j+\Delta E_j]$, measured in terms of PA and PF ($\Psi_j,\,\Pi_j$).
They are converted into the normalized Stokes parameters, i.e. the Stokes parameters divided by the total intensity ($I$), with a polar-Cartesian transformation:
\begin{equation}\label{eq:normalized_stokes_data}
    \left\{
    \begin{aligned}
        q_j &= \Pi_j\,\cos{\left(2\Psi_j\right)}\\
        u_j &= \Pi_j\,\sin{\left(2\Psi_j\right)}
    \end{aligned}
    \right.
\end{equation}

For the model space, we use a predefined energy grid, $E_i$, here defined as a logarithmic grid with 300 points from 1 to 2500\,keV.
From the 3ML model, we can compute the flux on this grid for each component, as $F^m_k(E_i)=F^m_{ki}$ (in photons\,s$^{-1}$\,cm$^{-2}$\,keV$^{-1}$). The polarization model is computed on the same energy grid, resulting in predicted PAs ($\Psi^m_{ki}$) and PFs ($\Pi^m_{ki}$) for each component. The Stokes parameters (in the same unit as the flux) are,
\begin{equation}\label{eq:stokes_comp}
    \left\{
    \begin{aligned}
        Q^m_{ki} &= F^m_{ki}\ \Pi^m_{ki}\,\cos{\left(2\Psi^m_{ki}\right)}\\
        U^m_{ki} &= F^m_{ki}\ \Pi^m_{ki}\,\sin{\left(2\Psi^m_{ki}\right)}
    \end{aligned}
    \right.
\end{equation}
We sum over the components in Stokes space to match the model and data energy bins,
\begin{equation}\label{eq:rebin_polar}
    \left\{
    \begin{aligned}
        I^m_j &= \sum_k B_{ij}\cdot F^m_{ki} \,\Delta E_i  \\
        Q^m_j &= \sum_{k} B_{ij}\cdot Q^m_{ki} \,\Delta E_i  \\
        U^m_j &= \sum_{k} B_{ij}\cdot U^m_{ki} \,\Delta E_i 
    \end{aligned}
\right.
\end{equation}
where $I^m$ is the integrated flux in units of photons\,s$^{-1}$\,cm$^{-2}$, found from the spectral model, and $B_{ij}$ is a re-binning matrix defined as,
\begin{equation}\label{eq:rebinning_matrix}
    B_{ij} = \left\{
    \begin{array}{ll}
        1 & \mbox{if } E_i \in \left[E_j, E_j+\Delta E_j\right].\\
        0 & \mbox{otherwise } \\
    \end{array}
\right.
\end{equation}
Finally, we normalized them by the integrated flux to find the predicted normalized Stokes parameters,
\begin{equation}\label{eq:normalized_stokes_comp}
    \left\{
    \begin{aligned}
    q^m_j &= Q^m_j/I^m_j\\
    u^m_j &= U^m_j/I^m_j
    \end{aligned}
    \right.
\end{equation}
These quantities can be compared directly with the data to find the optimal $\chi^2$ figure-of-merit. Optionally, the total polarization predicted by the model can be found with:
\begin{equation}\label{eq:stokes_to_polar_model}
\left\{
    \begin{aligned}
    \Psi^m_j &=\frac{1}{2} \arctan{\left(u^m_j/q^m_j\right)}\\
    \Pi^m_j &= \sqrt{\left(u_j^m\right)^2+\left(q_j^m\right)^2}
    \end{aligned}
    \right.
\end{equation}


\subsection{Polarization uncertainties}\label{sec:polar_uncertainty}

The uncertainties on the ($Q,U$) Stokes parameters are assumed equal, which is a reasonable approximation for most experiments \citep{Quinn_2012}. It follows that the uncertainties on the normalized Stokes parameters are also equal ($\sigma_{q_j}=\sigma_{u_j}\equiv\sigma_j$), and can be deduced from the uncertainties on the polarization.
This uncertainty is also approximately Gaussian, despite being the ratio of two quantities (Eq.\ref{eq:normalized_stokes_comp}), because the uncertainties on the intensities are negligible compared to the polarization, since the flux is calculated from the spectral model \citep{Quinn_2012}.
Using uncertainty propagation from the second relation of Eq.\ref{eq:stokes_to_polar_model}, we directly find that $\sigma_j=\sigma_{\Pi_j}$, without requiring the first relation on the PA. This simple equality can be understood from the fact that uncertainties are independent of the PA, due to our previous hypotheses.\\

\subsection{Modeling procedure}\label{sec:modeling}

The resulting $\chi^2$ is the sum of the spectral $\chi^2$ ($\chi_{spec}^2$), and polarization-dependent Stokes $\chi^2$ ($\chi_{pol}^2$) \citep{Kislat_2015}. Because of the previous hypotheses, mainly that the relative uncertainty of the intensity is much less than of the normalized Stokes parameters, we have $\sigma_Q\approx I\,\sigma_q$ and $\sigma_U\approx I\,\sigma_u$.
The residuals of the normalized Stokes parameters are therefore equal to the Stokes residuals, and we can write,
\begin{equation}\label{eq:chi2_data}
    \begin{aligned}
    \chi_{data}^2 &= \chi_{spec}^2 + \chi_{pol}^2 \\
        &= \sum_{s} \frac{(C_s-C^m_s)^2}{\sigma_s^2} + \sum_{j} \frac{\left( q_j-q^m_j \right)^2+\left( u_j-u^m_j \right)^2}{\sigma_j^2}
    \end{aligned}
\end{equation}
Some parameters are difficult to constrain, and instead of fixing them, we assumed a Gaussian prior for each of those parameters, $p_l$, with a mean $\mu_l$ and standard deviation $\sigma_l$,
\begin{equation}\label{eq:chi2_prior}
    \chi^2_{prior} =
        \sum_{l} \frac{(p_l-\mu_l)^2}{\sigma_l^2}
\end{equation}

The final chi-square used for the fit is then corrected by this prior chi-square: $\chi^2=\chi_{data}^2 + \chi^2_{prior}$.

We used the Levenberg-Marquardt least-square fitting algorithm implemented in the \texttt{lmfit} Python library to find the best parameters \citep{lmfit}.
To have a better constraint on some specific fit parameters, we computed the confidence intervals using the \textsc{conf\_interval} function of the same library.
This function fixes all the parameters except the one of interest, and minimizes the $\chi^2$ to find the confidence interval at 1-$\sigma$ of each parameter. This was applied for parameters with strong asymmetric uncertainties (see Section \ref{sec:result}). The fit statistics (reduced $\chi^2$ and p-value) are based on the $\chi_{data}^2$.\\

\begin{figure}[h!]
    \centering
    \includegraphics[width=\linewidth]{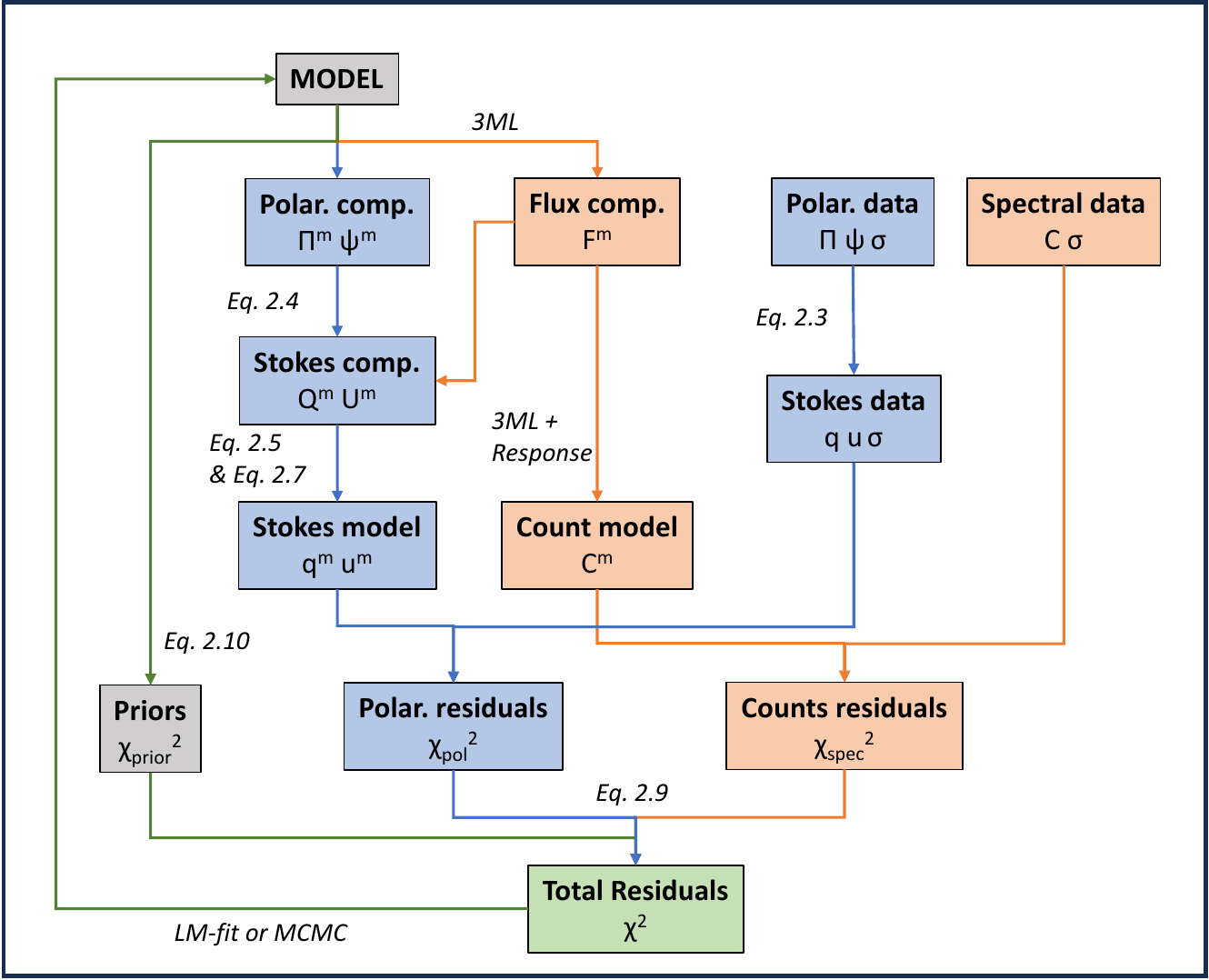}
    \caption{Workflow of the polarization fitting algorithm (Polar. = Polarization, Comp. = Components). Polarization quantities are in blue, and flux quantities in orange.}
    \label{fig:polar_fit_process}
\end{figure}

Finally, to assess the parameters posterior distributions and correlations in more detail, we also applied a Monte-Carlo Markov Chain (MCMC) algorithm, from the \texttt{emcee} Python library \citep{emcee}. The chain explores the log of the posterior distribution, so $-\frac{1}{2}\chi^2$ in our case. A summary of the fit algorithm's workflow is shown in Fig. \ref{fig:polar_fit_process}.

\section{Models}\label{sect:models}

We now discuss the model components used with the above described approach to model observational data. As there are very few data points for the polarization at high energies, we decided to confine ourself to models with only a limited number of free parameters.  We start with the IC emission, that usually cuts off around 100 \,keV \citep{Gierlinski_1997}, followed by the hard $\gamma$-tail that we describe as synchrotron emission. In the hard state, the thermal contribution of the accretion disk is negligible above 2\,keV, and so is, for mildly-absorbed sources such as \cygx which we focus on here, the interstellar absorption.
We did not include a reflection component, in order to keep the model as simple as possible while still capturing the main spectro-polarimetric features. In fact, for \cygx, the inclusion of reflection has only a small influence on the other IC parameters \citep{Dove_1998}.

\subsection{Inverse Comptonization}\label{sec:IC_model}

\subsubsection{Spectrum}

For the spectral model, we used the \texttt{comptt} model described in \cite{comptt_1994}. The main parameters are the optical depth, $\tau_p$, and electron temperature, $T_e$. The seed photon are represented by a Wien law with temperature $T_{seed}$, which can be interpreted as the accretion disk inner radius temperature. It can be constrained from the low-energy cut-off, which we do not observe since our data start at 2\,keV. We therefore fixed it to a value based on independent analysis \cite[$T_{seed}=0.2$\,keV, ][]{Wilms_2006}, as done previously \citep{Cangemi2021, Rodriguez_2015}.\\

\subsubsection{Polarization}\label{sec:ic_polar}

The IC polarization can be a non-trivial function of energy, depending on the geometry and opacity of the IC medium \citep{Schnittman_2010}. As we are mainly interested in understanding the hard-tail while using the X-ray polarization as an additional constraint, we used a phenomenological function to represent the evolution of PF with energy.
In the literature, linear models are common, (see e.g., \cite{Krawczynski_2022_cygnus}). However, this only works for the narrow energy ranges such as the one covered by IXPE, and in our case would diverge in the hard X-rays. In fact, the theory predicts that the PF rapidly saturates after multiple consecutive Compton up-scatterings \citep{Titarchuk_2025}. Above the thermal cut-off energy (above $\sim$100\,keV), the shape could become more complex.
Since the flux is dominated by the highly polarized hard-tail in this region, we did not attempt to include those variations, as they would not significantly influence the model.\\

To approximate this linear increase (seen in soft X-rays) and saturation (seen in hard X-rays) analytically, we used a log-sigmoid function, which has the properties of linearly increasing, and saturating near the energy $E_L$ at a maximum value $\Pi_M$. The IC PF, $\Pi_{IC}$, can be written as,
\begin{equation}\label{eq:logsigmoid_mod}
    \Pi_{IC}(E) = \Pi_M - l_1\cdot (\Pi_M-\Pi_1)\cdot \ln{\left [ \exp{\left(\frac{1-E/E_L}{s}\right)} +1 \right]},
\end{equation}
where $s$ is a smoothness parameter that controls the speed of the transition, and $l_1$ is a factor that depends only on $E_L$ and $s$.
The reference value of the PF, $\Pi_1$, is chosen at an energy of 1\,keV in order to give a universal, instrument-independent definition, following the same convention as spectrum normalization factors.
With the constraint of $\Pi_1\geq0$\%, this expression keeps the formula physically valid above 1\,keV.
Examples of log-sigmoid functions for different parameters are shown in Fig. \ref{fig:loglogistic_examples}. Physically, the saturating value $\Pi_M$ depends almost exclusively on the plasma opacity, Comptonized medium shape, and inclination. Due to the low statistics of the polarization data between 10--100\,keV where this saturation should be observed, we fixed this value using realistic parameters.
This is possible because the opacity is constrained mainly from the spectrum. Based on a spectrum-only fit (see Section \ref{sec:data}), we used an opacity of $\tau_p=1.1$ and an inclination of the inner disk of \ang{65;;} \citep{Krawczynski_2022_cygnus} to compute this saturating value.
For a plane-parallel, optically thick scattering medium associated with an accretion disk, this results in $\Pi_M\approx15$\% \citep{Sunyaev_1985}.\\

Depending on opacity, and assuming an axisymmetry of the corona around the jet axis, the expected PA is either parallel or perpendicular to the axis. The PA is close to constant in the soft X-rays \citep{Schnittman_2010}, and we therefore keep the model PA, $\Psi_{IC}$, constant with energy.\\


\subsection{Synchrotron from a cut-off power-law}

The hard-tail is highly polarized, and therefore assumed to be synchrotron emission from a non-thermal population of electrons. Generally, particle acceleration in astrophysical systems results in a power-law with a stretched exponential cut-off \citep{cutoff_shapes}. The Lorentz factors ($\gamma$) of the electrons follow:
\begin{equation}\label{eq:electron_distribution}
    N_e(\gamma)=K_e \gamma^{-p} \exp{\left[ - (\gamma/\gamma_{cut})^{\beta} \right]}
\end{equation}
where $\gamma_{cut}$ is the cut-off, $p$ is the power-law index for $\gamma\ll\gamma_{cut}$, $K_e$ a normalization factor, and $\beta$ a super-exponential index. $\beta$ and $\gamma_{cut}$ are determined by the underlying physical processes that accelerate and cool the electrons. Due to the low count statistics of the spectrum near 1\,MeV, we could not constrain the value of $\beta$ from the data.
Therefore, we fixed $\beta=2$ for the fits, which represents the case of Bohm diffusion and synchrotron cooling \cite{cutoff_index_bohm, Zdziarski_2014}.\\

\subsubsection{Spectrum}
The resulting photon spectrum from synchrotron emission also follows a cut-off power-law, with a different super-exponential index \citep{Lefa_2012},
\begin{equation}\label{eq:sync_photon}
    N(E)=K E^{-\Gamma} \exp{\left[ - \left(\frac{E}{g_{\beta}E_{cut}}\right)^{\delta}\, \right]}
\end{equation}
where $\delta=\beta/(\beta+2)$ is the super-exponential index, $\Gamma=\frac{p+1}{2}$ the photon index, $K$ a normalization factor (in ph\,s$^{-1}$\,cm$^{-2}$\,keV$^{-1}$), $E_{cut}$ the magnetic field ($B$) dependent photon cut-off energy, defined as,
\begin{equation}\label{eq:photon_cutoff}
    E_{cut}= \left(\frac{3\hbar e}{2m_ec}\right) B\gamma_{cut}^2,
\end{equation}
and $g_{\beta}$ a factor that corrects for the modified shape of the cut-off, defined as,
\begin{equation}
g_{\beta}=\frac{\beta}{2}\left( \frac{2}{\beta+2} \right)^{\frac{\beta+2}{\beta}}.
\end{equation}
For $\beta=2$, this factor is $1/4$, which modifies significantly the estimation of the cut-off energy.
Note that some authors also approximate the synchrotron emissivity with Dirac distribution approximation, which leads to an incorrect value of $\delta=\beta/2$ \citep{Lefa_2012}.\\

The value of $E_{cut}$ is obtained from the spectral fit, while the electron cut-off is deduced afterwards. Inverting Eq. \ref{eq:photon_cutoff}, we find,
\begin{equation}\label{eq:cutoff_electrons}
    \gamma_{cut} m_e c^2= 12.3\,\mathrm{GeV} \left(\frac{E_{cut}}{100\,\mathrm{keV}}\right)^{1/2} \left(\frac{B}{10^4\,\mathrm{G}}\right)^{-1/2},
\end{equation}
for expected conditions in BHXBs. The power-law index of the electrons is usually hard to constrain, as it requires a broad-band spectral analysis.
We therefore used a Gaussian prior for the photon index based on the SED analysis of \cite{Zdziarski_2014}, with a mean of $1.2$ and standard deviation of 0.1 (see Section \ref{sec:modeling}). This allows for the model to explore more parameter space while keeping a constraint from the radio to IR data.

\subsubsection{Polarization}\label{sec:polar_sync}

For a power-law distribution of electrons, the resulting PF is constant with energy \citep{Longair_2011}. In the case of an exponential cut-off, this holds only for $E\ll E_{cut}$, because the local power-law index will change near the cut-off. To compute this energy dependency explicitly, we need to integrate the electron population with the synchrotron kernels, using the variable of integration $x=E/E_c$, with $E_c=\left(\frac{3 \hbar e}{2m_e c}\right)\gamma^2B$ the critical synchrotron photon energy.
After performing the change of variable from electron to photon energy, the electron distribution transforms into a function $H$, where we dropped the $x$-independent factors that disappear in the following equations \citep[adapted from][]{cutoff_exp_polar},
\begin{equation}
    H(x,E)=x^{\frac{p-3}{2}} \exp{
    \left[
        -\left(\frac{x}{x_{cut}(E)} \right)^{-\beta/2}
    \right]},
\end{equation}
where we defined $x_{cut}=E/E_{cut}$.
If we had chosen a power-law electron distribution, $H$ would only be a function of $x$. Because we introduced a cut-off in the distribution, the photon energy $E$ is "coupled" with $x$, and the function $H$ retains an explicit dependency with $E$ that cannot be removed by a change of variable.
The PF is then calculated as the ratio \citep{Longair_2011},
\begin{equation}\label{eq:sync_polar}
    \Pi(E)=\frac{\int_{0}^{\infty}G(x)H(x,E)\,dx}{\int_{0}^{\infty}F(x)H(x,E)\,dx}
\end{equation}
where the kernel functions are \citep{Westfold_1959},
\begin{equation}\label{eq:kernel_sync}
    \left\{
    \begin{aligned}
        F(x)&=x\int_x^{\infty}K_{5/3}(u)du\\
        G(x)&=x K_{2/3}(x)
    \end{aligned}
    \right.
\end{equation}
with $K_{\alpha}$ the modified Bessel function of the first kind of order $\alpha$.
For a power-law ($x_{cut} \rightarrow \infty$), $H$ only depends on $x$, and we recover the classic result of energy-independent polarization. As a general rule, if the variables $x$ and $E$ in the electron distribution can be separated ($N(E,x)=f(E)\times g(x)$), as in the power-law case, then the polarization is constant with energy, due to the cancellation of $f(E)$ in Eq.\ref{eq:sync_polar}. Otherwise, it depends on $E$, as in our model. The full numerical calculation is explained in Appendix \ref{sec:grid}.\\

\begin{figure}[h]
    \centering
    \includegraphics[width=\linewidth]{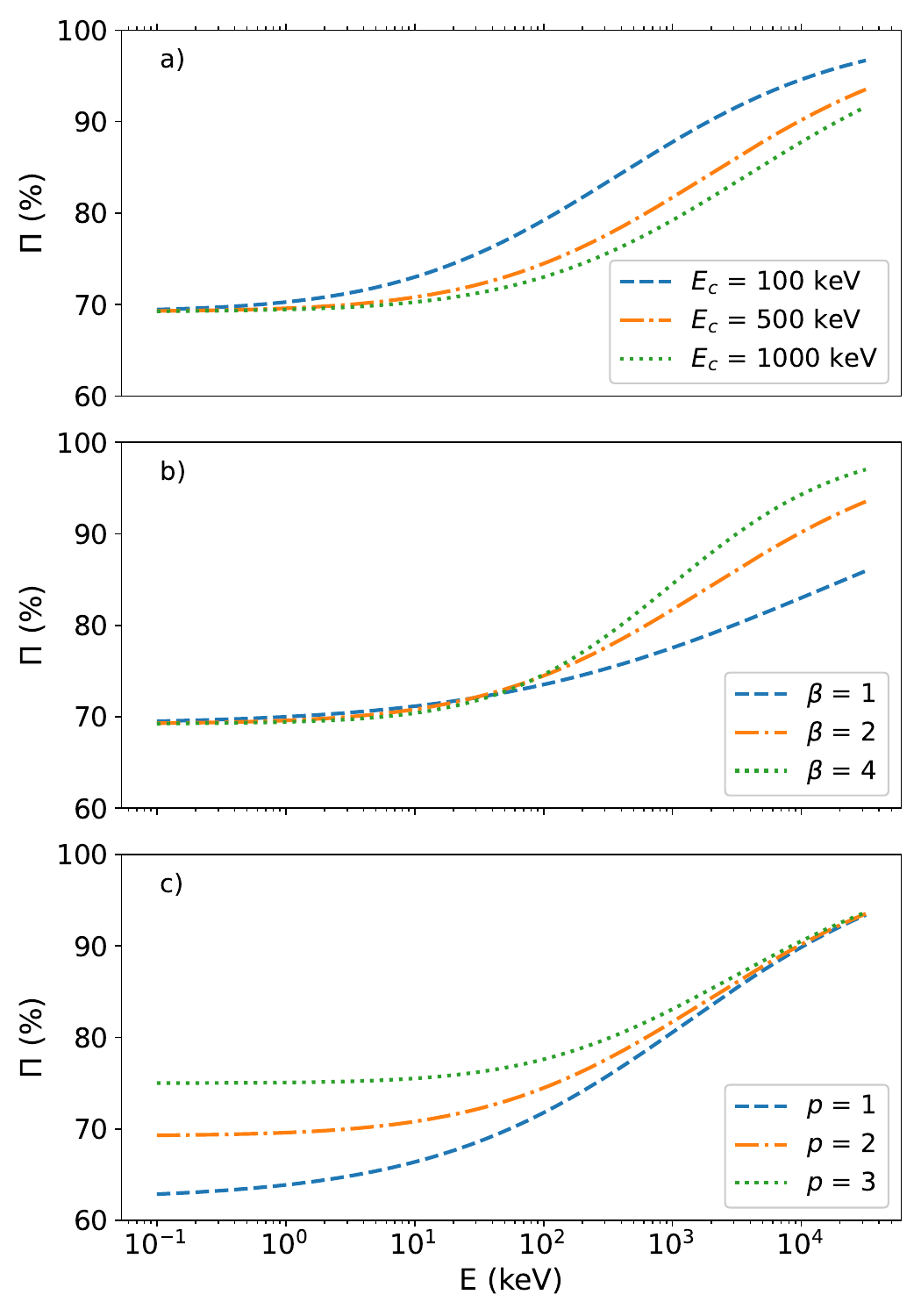}
    \caption{Polarization fraction of synchrotron emission from a stretched cut-off power-law electron distribution. a) For different values of $E_{cut}$. b) For different values of $\beta$. c) For different values of $p$. Default values for all plots are $E_{cut}=500$ keV, $p$=2, $\beta$=2.}
    \label{fig:cutoff_polar_test}
\end{figure}

Examples with different values of $E_{cut}$, $\beta$ and $p$ are shown in Fig. \ref{fig:cutoff_polar_test}. $E_{cut}$ controls the position at which the polarization starts to significantly deviates from the classic power-law result. $\beta$ impacts the speed of the transition. $p$ only influences the polarization in the power-law region, prior to the cut-off, as expected. The synchrotron PA, $\Psi_{sync}$, is constant with energy and perpendicular to the magnetic field. If the magnetic field is not perfectly ordered, the resulting PF can be lower by an energy-independent factor $0\leq f\leq1$ \citep{Russell_2014}. In our case, the uncertainty on the PF above 100\,keV are too high to constrain this value. Since there are no reference value in the literature, we fixed it at 1, corresponding to the case of a highly ordered magnetic field, which is the favored geometry \citep{Zdziarski_2014}.\\

All the models used for polarization and spectrum with their associated parameters are summarized in Table \ref{tab:summary_models}. We note that for the hard synchrotron tail, all the parameters except $\Psi_{sync}$ have a direct impact on both the spectrum and polarization, confirming the need for a combined spectro-polarization analysis. \\

\begin{table}[h!]
\centering
\caption{Summary of the models parameters.}
\label{tab:summary_models}
{\renewcommand{\arraystretch}{1.3}
\begin{tabular}{c|cc} 
Component & Parameters & Description \\
\hline
\emph{IC}    & $kT_{seed}$    &   Seed photon energy  \\
    & $kT_e$    &  Thermal electrons energy \\
    & $\tau_p$    &   Plasma optical depth  \\
    & $K_{IC}$    &   IC normalization \\
    & $\Psi_{IC}$  &  IC PA   \\
    & $s$    &  PF shape parameter  \\
    & $\Pi_1$    &   IC PF at 1\,keV   \\
    & $\Pi_M$    &    IC PF saturation \\
    & $E_L$    &   Transition energy   \\
    \hline
    
\emph{Synchrotron}    & $\Gamma_{tail}$    &  Photons power-law slope  \\
    & $E_{cut}$  & Photon cut-off energy \\
    & $\beta$  & Electron super-exponential \\
    & &  index \\
    &  $K_{sync}$ &    Synchrotron normalization  \\
    & $\Psi_{sync}$   &  Synchrotron PA  \\
    \hline
    \end{tabular}
    }
\end{table}

\section{Data}\label{sec:data}
For our analysis, we use the average broad band spectrum of \cygx over more than a decade as well as a combination of polarization measurements from different instruments to cover different energy bands.
We focused on the hard state of the source, in which the presence of the compact jet has already been established \citep{Stirling_2001, Rahoui_2011}. The hard X-ray tail is also most prominent in this state \citep{Cangemi2021}, allowing us to investigate it in detail using our synchrotron model.

\subsection{Spectral data}

Our spectral model is basic, and therefore does not require extensive spectral data.
For \cygx, a coherent analysis of the INTEGRAL spectral data was already performed by \cite{Cangemi2021}, in particular in the HS. This analysis conveniently covers the entire energy range required to fit our model parameters. It includes the period from 2003 June 9$^{th}$ to 2017 November 7$^{th}$, which is also in the same era as most polarization data points, in particular for SPI and IBIS.
Overall, this gives a coherent spectrum, with the same observation periods for all the instruments.\\

The HS observations were selected following the classification scheme of \cite{Grinberg_2013}. The data are from the JEM-X (3--20\,keV), ISGRI (30--400\,keV), and SPI (30--1500\,keV) instruments, covering the soft X-ray to soft $\gamma$-ray range. In the original publication, there are two spectra for the ISGRI instrument, due to the different pipeline versions covering different era and we only used the OSA11-version spectrum. A normalization constant is used to account for systematic errors. The constant of SPI is fixed at 1, as this instrument is considered the most thoroughly calibrated of the three \citep{Diehl_2018, Jourdain_2020}. A fit using these spectral data alone ($\chi^2=\chi^2_{spec}$) allows us to constrain $\tau_c=1.1$ and fix the value of $\Pi_M$, as explained in Section \ref{sec:ic_polar}.

\subsection{Polarization data}

\begin{figure}[h!]
    \centering
    \includegraphics[width=\linewidth]{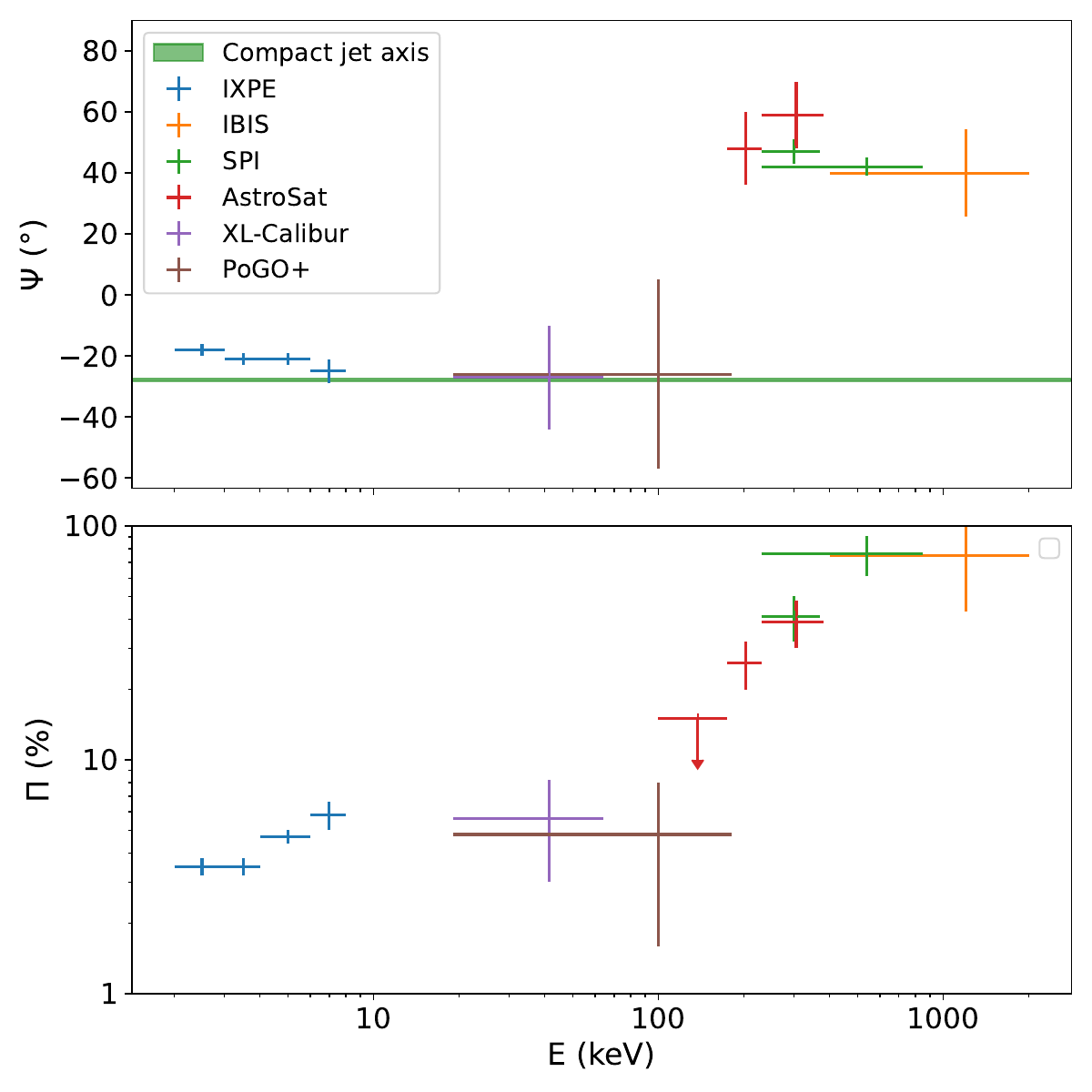}
    \caption{Polarization parameters of \cygx measured by different instruments in the HS.}
    \label{fig:sped_data_no_model}
\end{figure}

For \cygx, both the soft and hard X-rays polarization have been measured and confirmed by independent missions. In this persistent source, the compact jet axis was also resolved, and measured at a projected angle on the sky of \ang{-25;;} \citep{cygnus_compact_jet}. We summarize all HS measurements found in the literature from the past 25 years in Table \ref{tab:cygx1_polar_other} and Fig. \ref{fig:sped_data_no_model}.\\

\begin{table*}[h!]
\centering
\begin{threeparttable}
\caption{Summary of 21$^{st}$ century high-energy polarization measurements of \cygx in the Hard State. Confidences intervals are given at 1-$\sigma$.}
\label{tab:cygx1_polar_other}    
{\renewcommand{\arraystretch}{1.3} 
\begin{tabular}{cccccc}
Instrument & Years & E (keV) & $\Psi$ (°) & $\Pi$ (\%) & Ref. \\
\hline
INTEGRAL/IBIS & \emph{2003--2010} & 400 -- 2000 & 40 $\pm$ 14 & 75 $\pm$ 32 & \cite{Rodriguez_2015}\\
INTEGRAL/SPI & \emph{2003--2009} & 230 -- 370 & 47 $\pm$ 4 & 41 $\pm$ 9 & \cite{Jourdain_2012_polar}\\
&& 230 -- 850 & 42 $\pm$ 3 & 76 $\pm$ 15 & \\
AstroSat/CZTI & \emph{2019} & 100 -- 380 &  56 $\pm$ 11 & 23 $\pm$ 4 & \cite{Chattopadhyay_2024} \\
&& 100 -- 175 &  - & <15 & \\
&& 175 -- 230 &  48 $\pm$ 12 & 26 $\pm$ 6 & \\
&& 230 -- 380 &  59 $\pm$ 11 & 39 $\pm$ 9 & \\
PoGO+ & \emph{2016} & 19 -- 181 & 154 $\pm$ 22  & 4.8 $\pm$ 3.7 & \cite{Chauvin_2018} \\
XL-Calibur & \emph{2024} & 19 -- 64 &  -26 $\pm$ 31 & 5.6 $\pm$ 2.6  & \cite{xlcalibur_2025} \\
IXPE & \emph{2022} & 2 -- 8 &  -21 $\pm$ 1  &  4.0 $\pm$ 0.2   & \cite{Krawczynski_2022_cygnus}\\
&& 2 -- 3 &  -18 $\pm$ 2 & 3.5 $\pm$ 0.3   & \\
&& 3 -- 4 &  -21 $\pm$ 2 & 3.5 $\pm$ 0.3   & \\
&& 4 -- 6 &  -21 $\pm$ 2 & 4.7 $\pm$ 0.3   & \\
&& 6 -- 8 &  -25 $\pm$ 4 & 5.8 $\pm$ 0.8   & \\

\hline
    \end{tabular}}
    \end{threeparttable}
\end{table*}

We did not consider OSO-8 observations as IXPE ones are much more recent and closer in time to other data points. In their study of \cygx using AstroSat/CZTI observations, \cite{Chattopadhyay_2024} also separate the HS into pure HS (PHS) and hard intermediate state (HIMS). Their closest measurement in time to our spectral data is the HIMS, performed in 2019, and shows strikingly similar PF to SPI in their overlapping energy band (see Table \ref{tab:cygx1_polar_other}).
The lower-limit of SPI found by \cite{Jourdain_2012_polar} is also discarded, as our inference procedure does not take into account survival probability fitting.\\

A clear distinction between the high PF/jet-misaligned PA in the soft $\gamma$-rays regime and the low PF/jet-aligned PA of the soft X-rays can be seen in Fig. \ref{fig:sped_data_no_model}. In the hard X-rays (10--100\,keV), polarization measurements have a low significance (p-value>1\%), and we therefore used a slightly different approach to what is described in Section \ref{sec:polar_uncertainty}.
For XL-Calibur, the authors directly give the values of $q$ and $u$ with errors and we use those \citep{xlcalibur_2025}.
For PoGO+, the original paper by \cite{Chauvin_2018} does not report Stokes parameters, but a later article gives a 90\% confidence contour around a 2D-Gaussian in the normalized Stokes space ($q,u$) with a radius of 6.9\% \citep{Chauvin_2019}. By converting this value to a 1D 68\% confidence interval, assuming $\sigma_u=\sigma_q$, we find $\sigma=3.2$\%.\\


Some polarization measurements were taken after 2017, which could lead to some discrepancies if the spectrum or polarization changes. We note however that polarization measurements made with INTEGRAL \citep{Rodriguez_2015, Jourdain_2012_polar} cover large time periods of several years, and yet amount to high PF ($\Pi\gtrsim60\%$). Had polarization (angle or fraction) been strongly time-dependent, the resulting integrated polarization would have been probably lower, as explain in Section \ref{subsec:polar_mixing_example}. Overall, the high polarization on long integrated observations confirms the stability of the source polarization within the HS, justifying the use of non-simultaneous observations in hard X-rays.\\

For the soft X-rays, \cite{Kravtsov_2025} showed that the PF is correlated with the hardness of the spectrum (Fig. 4 of their article). We note, however, that those variations are quite small, within $\pm$1\% in the HS, and therefore acceptable for our empirical model of IC polarization.
The same authors showed that the polarization also varies with the orbit of the binary. In our data set, the polarization measurements of IBIS, SPI, PoGO+ and IXPE are all longer than the orbital period (5.6 days). Although the XL-Calibur and AstroSat measurements were slightly shorter than one orbit (5 and 4 days respectively), the polarization at these energies was shown to not vary significantly with time \citep{Chauvin_2018}.
Our study is therefore performed on orbital phase integrated observations.

\section{Results}\label{sec:result}

\begin{figure}[h]
    \centering
    \includegraphics[width=\linewidth]{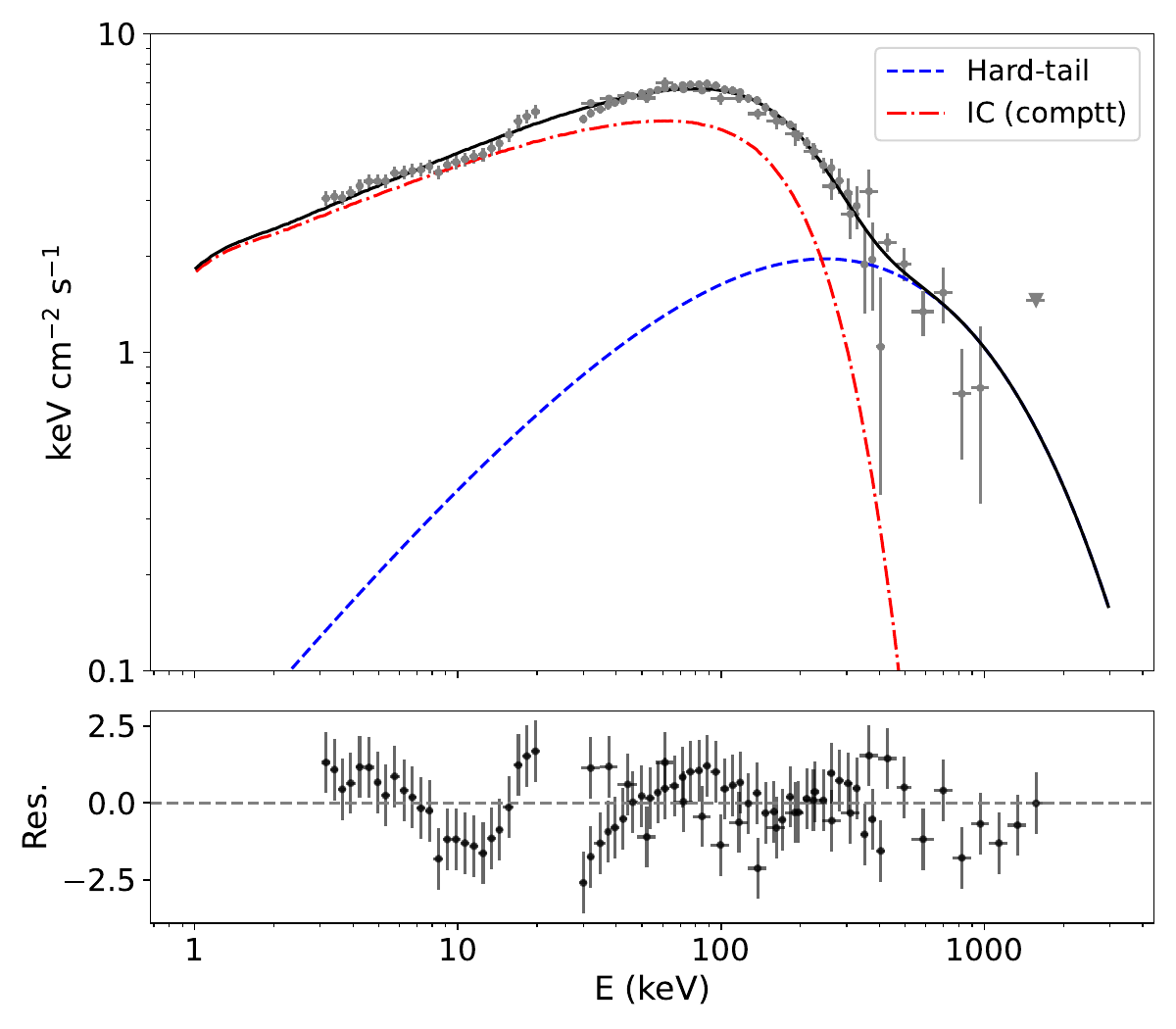}
    \caption{Spectral fit of \cygx INTEGRAL data (grey) with each component shown in blue (synchrotron) and red (IC), and total model in black. Associated residuals are shown in the lower panel.}
    \label{fig:spectral_fit_sync_exp}
\end{figure}

\begin{figure}[h]
    \centering
    \includegraphics[width=\linewidth]{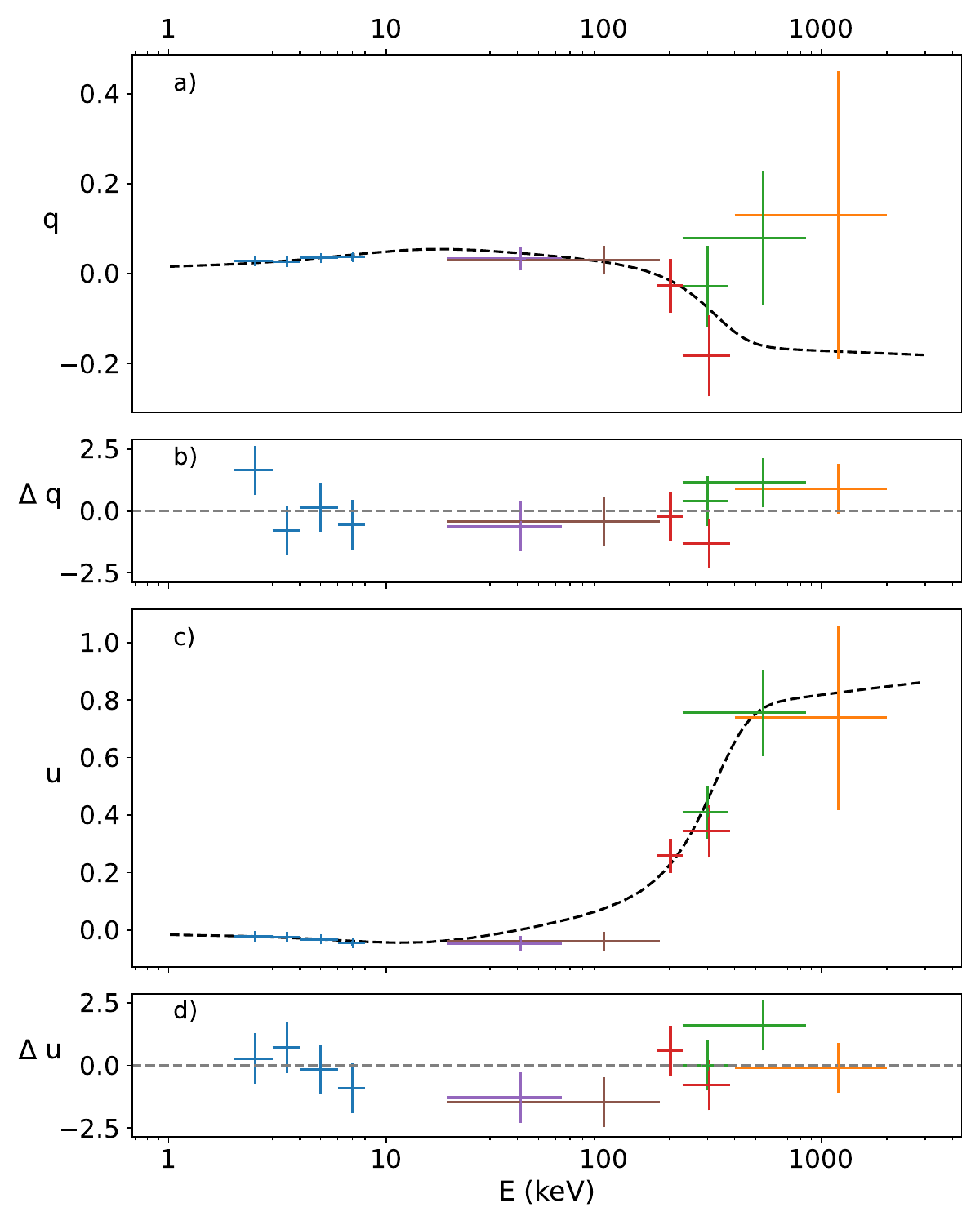}
    \caption{Best fit model (black dashed) of the polarization-dependent Stokes parameters of \cygx (panels a and c), with associated residuals (panels b and d).}
    \label{fig:polar_stokes_res_sync_exp}
\end{figure}

We applied our spectro-polarization method presented in Section \ref{sec:method} to model the data of \cygx discussed in the previous section, using the model consisting of IC and synchrotron from a cut-off power-law electrons distribution, as discussed in Sect.\,\ref{sect:models}. Accounting for the instrument normalizations, there are 11 free parameters. The best fit parameters for both components are shown in Table \ref{tab:best_fit_parameters}, along with the fit statistics. The fit is performed on the Stokes parameters, which are shown along the model in Fig.\ref{fig:polar_stokes_res_sync_exp}. We obtain a reduced $\chi^2$ of 1.05 for 97 degrees of freedom, giving a $\chi^2$ p-value of 35\%, confirming that the data is well represented by our model. 
The reduced $\chi^2$ cannot be defined for the spectrum and polarization separately, since they share common parameters. Nevertheless, the ratios of $\chi^2/N_{bins}$ for the spectrum and polarization (1.0 and 0.8, respectively) roughly indicate their contribution to the residuals.
The spectrum seems to contribute more to the total residuals per data bin. Adding another spectral component such as reflection of the IC could improve this balance.
The resulting spectral fit is shown in Fig.\ref{fig:spectral_fit_sync_exp} with the residuals. The synchrotron component crosses over the IC at an energy of 245\,keV, and decays following a stretched exponential $\sim \exp{\left(y^{1/2}\right)}$ (for $\beta=2$ and $y=E/E_{cut}$).\\

\begin{table}[h!]
\centering
\caption{Best fit parameters with 1-$\sigma$ confidence intervals, for the IC, synchrotron components, and instrument normalization. * indicates the parameter was fixed. Fit statistics are shown at the bottom.}
\label{tab:best_fit_parameters}
{\renewcommand{\arraystretch}{1.3}
\begin{tabular}{c|cc} 
 & Parameter & Best fit value \\
\hline
\emph{IC}   & $kT_{seed}$ & 0.2 keV* \\
 & $kT_e$ & $44.7_{-1.7}^{+1.9}$ keV \\
 & $\tau_p$ & $1.10\pm0.06$ \\
 & $\Psi_{IC}$ & $-30.3\pm2.4$ ° \\
 & $s$ & 1* \\
 & $\Pi_1$ & $4.2^{+1.0}_{-0.8}$\% \\
 & $\Pi_M$ & 16\% * \\
 & $E_L$ & $4.4_{-0.8}^{+1.0}$ keV \\
 \hline
\emph{Sync.}  & $\Gamma_{tail}$ & $1.06 \pm 0.09$ \\
 & $\beta$ & 2 * \\
 & $E_{cut}$ & $283_{-58}^{+83}$ keV \\
 & $\Psi_{sync}$ & $50.4\pm3.2$ ° \\
 \hline
\emph{Inst. Norm.}  & $C_{SPI}$ & 1 * \\
 & $C_{JEMX}$ & $0.88\pm0.02$ \\
 & $C_{ISGRI}$ & $1.00\pm0.01$ \\
 \hhline{=|==}
 \emph{Stat.} & $\chi_r^2$ & 1.06 (103.9 / 98) \\
& $\chi_{spec}^2$ & 85.1 (87 bins) \\
& $\chi_{pol}^2$ & 16.7 (22 bins) \\
& $\chi_{prior}^2$ & 2.1 (1 prior) \\
\hline
    \end{tabular}
    }
\end{table}

\begin{figure}[h]
    \centering
    \includegraphics[width=\linewidth]{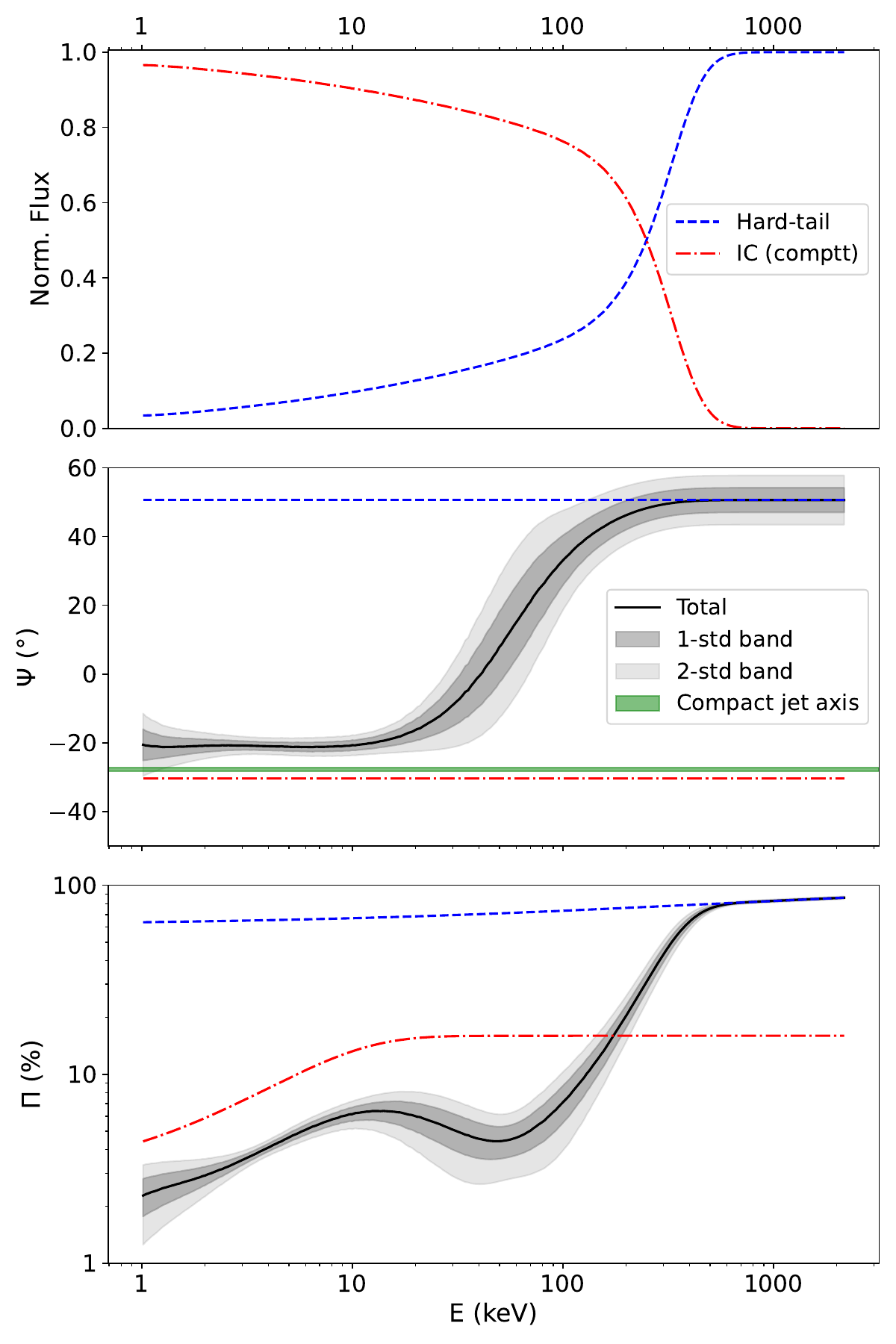}
    \caption{Polarization angle and fraction of the total model (black). Individual component are shown in blue (Synchrotron) and red (IC).}
    \label{fig:polar_model_sync_exp_log}
\end{figure}

To explore the parameter distributions in more detail, we ran a MCMC algorithm. We used 100 walkers for 12 free parameters. The mean autocorrelation time of the chains is $\approx580$, so we used 50,000 steps and burned the first 1,500 steps in order to obtain statistically sound sample of the parameter space \citep{emcee}. We show the 2D correlations of the main parameters of interest in Fig. \ref{fig:emcee_corner_best}, along with the 1D distribution. Although most most parameters are found largely uncorrelated and mostly symmetric, we found three pairs of parameter which show significant correlation. 
For the opacity and electron temperature, the correlation is understandable as both parameters influence the power-law slope of the IC \citep{Sunyaev_1980}.
The correlation parameter between $E_{cut}$ and $\Gamma_{sync}$ is also important, with a positive correlation parameter of $+0.84$. Indeed, a harder index (lower $\Gamma$ value) increases the flux at high-energy, which can be compensated by an earlier cut-off.
The correlation between the two PAs is more surprising. This could arise due to the mixing in the soft X-rays, where a lower $\Psi_{IC}$ can be compensated by a higher $\Psi_{sync}$, resulting in a correlation.
Including finer spectra above 1\,MeV would likely remove both of these correlations.
The $E_L$ parameter -- governing the transition from linear to saturated -- has a tail distribution with a bias towards high energy. This could be due to the lack of precise measurements between 10--100\,keV. Overall, the confidence intervals are close to the estimation from the least-squares method.\\

Finally, since the Stokes parameters are hard to interpret physically, we show the resulting model in Fig.\ref{fig:polar_model_sync_exp_log} for the polarization angle and fraction, as well as the normalized flux of each component ($F^m_{ki}/F^m_i$).
We used the last 10,000 steps from the MCMC run in order to find the standard deviation of our model in each energy bin. 
This allows us to show a band with 1 and 2 standard deviation in the plots (light grey).
A clear transition is seen between the IC and hard-tail angles, near $\sim$100\,keV, when synchrotron becomes dominant. The PF shows a peculiar behavior with a peak near 10\,keV and a local minimum near 40\,keV, where both components have similar contributions (same $\Pi(E)\times F(E)$) that cancel out because of their nearly orthogonal PAs (see Fig.\ref{fig:pa_mixing_example}). The PF then saturates at its maximum value near 500\,keV, where the synchrotron component dominates the spectrum. The presence of the cut-off in this energy range also allows for very high PF, much above the power-law limit (see Section \ref{sec:polar_sync}).\\

\section{Interpretation}\label{sec:interpretation}

The best fit parameters are interpreted to give a better understanding of both physical processes at play. We focus mainly on the synchrotron hard-tail, while the IC is discussed briefly.

\subsection{Spectrum}\label{sec:interpret_spec}
The thermal electron temperature ($kT_e=\kte^{+\kteplus}_{-\kteminus}$\,keV) and opacity ($\tau_p = \taup \pm \tauperr$) are both tightly constrained and similar to other studies \citep{Russell_2014, Cangemi2021}. The most interesting parameter that we were able to constrain is the photon cut-off energy, $(\Ecut^{+\Eplus}_{-\Eminus}) \times 10^{2}\ \mathrm{keV}$, which is tightly linked with the non-thermal electron distribution.
\cite{Russell_2014} used a slightly lower value of 330\,keV, although it is not clear whether the value was found from their fit or if it was fixed. This difference is also due to the spectral cut-off we used (see Eq. \ref{eq:sync_photon}), which differs from what is commonly used in the literature \citep[$\delta=1$, from the Dirac distribution approximation][]{Zdziarski_2014}.
An important caveat is the absence of reflection, which can potentially affect the spectral shape, and indirectly influence the inferred synchrotron cut-off.
In \cite{Jourdain_2012_spec}, the hard-tail was also modeled with a cut-off powerlaw, but was interpreted as a secondary IC emission from thermal electrons of temperature $kT_e=123\pm10$\,keV, which is not compatible with the polarization measured.\\

To link this value with the original electron distribution, we use Eq.\ref{eq:cutoff_electrons} to find,
\begin{equation}
    \gamma_{cut} m_e c^2= \elecgev^{+\elecgevplus}_{-\elecgevminus}\,\mathrm{GeV} \left(\frac{B}{10^4\,\mathrm{G}}\right)^{-1/2},
\end{equation}
for a typical ${10^4\,\mathrm{G}}$ magnetic field strength.\\

Assuming Bohm diffusion, \cite{Markoff_2001, Markoff_2003} made an estimation of this maximum energy based on the synchrotron cooling time and acceleration time,
\begin{equation}
    E_{cut}\approx 500\,\mathrm{keV}\left(\frac{n}{100}\right)^{-1} \left(\frac{u_{sh}}{c} \right)^2,
\end{equation}
where $n$ measures how many times the particle gyrates per shock crossing and $u_{sh}$ is the shock speed. For a typical estimated value of $n$ between 10 and 100 \cite{Jokipii_1987}, the cut-off we find implies $u_{sh}\approx (\ulow-\uhigh)\,c$, i.e. a Lorentz factor of the shock of $\Gamma_{sh}\approx \gammashlow - \gammashhigh$. This range of values is compatible with the expected Lorentz factor of compact jets in the HS \citep[$\Gamma\leq2$][]{Fender_2009}.\\

The photon index of the hard-tail is also constrained, although the Gaussian prior has a strong influence. For synchrotron emission, the electron index is $p=1.12 \pm 0.18$, much harder than in previous studies \citep{Zdziarski_2014}. Magnetic reconnection can produce hard power-laws with index reaching $p=1$ in the limit of high magnetization parameter \citep{reconnection_2014}, making this acceleration process favored to explain the non-thermal electron population.\\

\subsection{Synchrotron polarization angle}\label{sec:sync_pa}

The jet axis at launch is tightly constrained from radio observations of the compact jet modeled throughout the binary orbit, at $\theta_{jet}=-27.8 \pm 0.5$° \citep[model 2][]{Prabu_2026}.
The synchrotron PA is then $\Psi_{sync}-\theta_{jet}=77\pm3$° away from the compact jet axis, which is more than 4 standard deviation away from 90°. This is in contradiction with previous studies which hypothesized the soft $\gamma$-rays PA to be perpendicular to the jet \citep{xlcalibur_2025}, i.e. the magnetic field being parallel to the jet. From our model, the magnetic field seems to be in a less trivial configuration. To explain the discrepancy, one solution is that the pitch angle of the non-thermal electrons at the base of the jet could be anisotropic, which would induce a change in PA, especially in the cut-off region of the spectrum \citep{Bjornsson_1985, Russell_2014}.\\

We propose an alternative hypothesis that does not require a non-axisymmetric configuration, unlike previous explanations. In the presence of a moderate jet opening angle and helical magnetic field lines, the PA can in fact become misaligned with the expected directions (perpendicular or parallel). Helical magnetic fields are commonly seen in active galactic nuclei from observation of optically thin synchrotron emission with VLBI technique near the "jet-launching" regions \citep{helical_agn}. Theoretically, magnetic fields twisted into helical shape are also expected near the spinning BH in BHXB, due to frame dragging \citep{blandford_znajek, Tchekhovskoy_2011}.\\

\begin{figure}
    \centering
    \includegraphics[width=\linewidth]{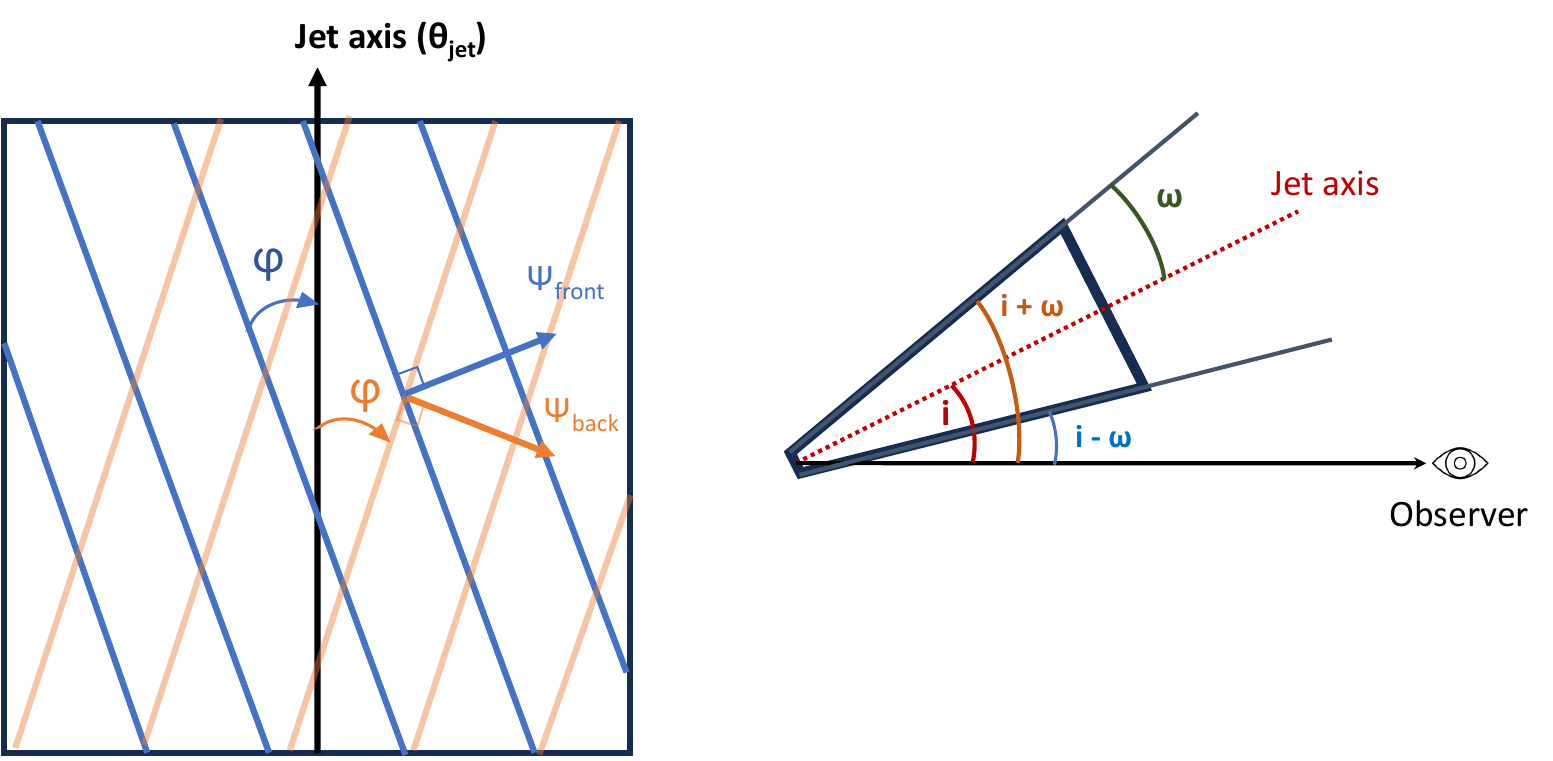}
    \caption{Geometric sketch of the helicity boosting scenario. Left: a zoom on a vertical slice of the jet seen from the side, with the front magnetic field shown with blue lines, whereas the back magnetic field is shown in orange. Right: Inclination and viewing angle of the jet with respect to an observer.}
    \label{fig:helicity_boosting}
\end{figure}

In the case of \cygx, the magnetic field would need to be nearly parallel to the jet, with a helicity pitch angle of $\varphi$,  defined as the angle between the magnetic field direction and the jet axis, in the region producing soft $\gamma$-rays (see Fig. \ref{fig:helicity_boosting}, left). In the frame of the jet, viewed from the side, the front layer of the jet produces a PA of $\Psi_{front}\approx\theta_j +\pi/2- \varphi$ (in blue), while by axisymmetry, the back layer has the opposite pitch angle, $\Psi_{back}\approx\theta_j +\pi/2+ \varphi$ (in orange). In the absence of any other effect, this results in a total PA perpendicular to the jet (see Eq.\ref{eq:mixing_2_component}). However, if the jet has a half-opening angle $\omega$, the front part has a viewing angle ($i-\omega$) with the observer, while the back has a wider angle ($i+\omega$) (see Fig. \ref{fig:helicity_boosting}, right). The front part will therefore experience a stronger apparent Doppler boosting towards the observer, similarly to a GRB \citep{grb_doppler}, as well as a different projection angle. Based on theoretical models and simulations, the opening angle is thought to be much higher close to the BH, where $\gamma$-rays are likely produced, than what is inferred from radio observations at larger scales \citep{Chatterjee_2019}.
For a wide enough opening angle ($\omega\gtrsim10^{\circ}$), and moderate Lorentz factor -- such as the one found in the previous section -- the front layer can dominate the emission, and hence the polarization. Since the front polarization dominates, the misalignment, $|\Psi_{sync} - \pi/2-\theta_j| \approx \helicity^{\circ}$, could then be used to infer the helicity parameter.
For any inclination $i$, this results in an upper-limit on the intrinsic helicity of $\approx \helicity ^{\circ}$.
Determining a more precise value would involve much more calculation, and is outside the scope of this paper.\\

This shifted PA effect therefore requires three hypothesis: large opening angle, Doppler boosting, and helical magnetic field.
Although each of those three hypothesis have been proposed separately before \citep{Russell_2014, Rodriguez_2015}, only their combination leads to the observed PA.
At higher wavelength, the optically thin synchrotron PA is parallel to the jet, i.e. produced in a different region where the magnetic field is toroidal \citep{Russell_2014}.
Overall, this explanation is more satisfying as it does not involve a break of the axisymmetry near the jet, which would imply the existence of a preferred direction (other than the jet axis).

\subsection{BH spin and helicity}

In the helical magnetic field scenario, if we assume that the emission comes from the approaching jet, then the sign of $\varphi$ also gives an additional information. Since $\varphi>0^{\circ}$, this means that the magnetic field is winding clock-wise when viewed from the top (left-handed helix). During the Blanford-Znajek process, the originally poloidal magnetic field lines twist in the direction opposite to the product of BH angular speed ($\Omega_{BH}$) and poloidal magnetic field ($B_{p}$) \citep{Komissarov_2004},
\begin{equation}
    B_{\phi}\propto -(\Omega_{BH}-\Omega_{F}) B_{p},
\end{equation}
where $\Omega_F<\Omega_{BH}$ is the angular speed of the magnetic field lines. 
In the case of \cygx, the sign of $\varphi$ implies $B_{\phi}<0$.
We deduce that the BH spin and poloidal component of the magnetic field near the BH have the same direction, although we cannot know the actual direction of either of them.
For instance, if the spin is oriented upward along the approaching jet, the magnetic field should be directed upward as well.
We note that this degeneracy could potentially be broken if the circular polarization is somehow measured.\\

\subsection{IC polarization}

Fig. \ref{fig:polar_model_sync_exp_log} shows a clear gap between the total expected PF in soft X-rays (black) and the IC PF alone (red), which is due to the contamination of the hard-tail (blue).
The IC polarization parameters are hard to interpret directly since this model is purely phenomenological.
However, the IC model in the 2--8\,keV region can be approximated as a line with an offset of $A=(6.7\pm0.1)\%$ at 1\,keV and a slope of $B=(1.05\pm0.03)\%/$keV. Those values are quite different from those obtained when fitting the IXPE data alone with a single line component \citep[$A=(2.9 \pm 0.4)$\% and $B = (0.58 \pm 0.15)\%/$keV,][]{Krawczynski_2022_cygnus}. This clearly shows that despite being only dominant in soft $\gamma$-rays, the hard-tail can fundamentally influence the shape in the soft X-rays and potentially change the interpretation of the IC polarization significantly.\\

The IC PA is $2.5\pm2.3$° away from the compact jet axis, i.e. compatible with the jet direction within 2 standard deviation (see Fig.\ref{fig:polar_model_sync_exp_log}, middle).
The IXPE PA in the full 2--8\,keV energy range is $-7\pm1$° away from the jet (see Table \ref{tab:cygx1_polar_other}), which is significantly further away.
Similarly as the PF, this difference is due to the addition of the hard-tail in our model, which influences the soft X-rays and requires a lower $\Psi_{IC}$ than what would be expected from a single IC component.
We therefore demonstrated with our model that the difference between the compact jet and IC PA is not intrinsic to the source geometry, but instead due to the overlap of different emission component in the spectrum.
Interestingly, the soft X-rays PA in the soft state is $-25.7 \pm 1.8$°, i.e. $2.1 \pm 1.9$° away from the jet axis, which is significantly closer than in the hard state \citep{Steiner_2024}. This corroborates our conclusion, since the soft state is unlikely to be contaminated by the strongly polarized hard-tail, as was shown by the polarization upper-limit in the 100--380\,keV range \citep[$\Pi<12\%$ at 99\% confidence level,][]{Chattopadhyay_2024}.\\

Overall, the close alignment between the IC PA and the compact jet axis disfavors the warped-disk interpretation, which requires an inclination of at least $45^\circ$ \citep{Krawczynski_2022_cygnus}, compared with the $27^\circ$ inclination inferred for the compact jet \citep{Prabu_2026}. If the two axes were misaligned by this difference in inclination, we would expect the IC PA and the projected position angle of the compact jet to differ by a comparable amount of at least $\sim22^\circ$. Instead, our model finds that the two are closely aligned. If the warped-disk interpretation is correct, this alignment would therefore require the disk warp to be oriented almost directly along our line of sight. While such a configuration is geometrically possible, it requires a specific fine-tuning of the warp orientation, making it less natural under the principle of Occam's razor. A more likely explanation relies on a relativistic outflow in a conical sheath around the jet with a large opening angle $\sim 30$° and bulk Lorentz factor of $\gtrsim 1.5$ \citep{Jason_2024}. Interestingly, those hypotheses are consistent with our explanation of the shifted synchrotron PA (see Section \ref{sec:sync_pa}), and our derived value of $\Gamma_{sh}$ (see Section \ref{sec:interpret_spec}), which potentially hints towards a common medium for both components in the spectrum.

\subsection{Polarization predictions}

Some polarization predictions from our model are shown in Table \ref{tab:models_prediction}, using the effective polarization energy ranges of different future missions. This includes XPoSat \citep[][active]{xposat_paper}, COSI \citep[][accepted]{cosi_main}, and the three proposed instruments: PHEMTO \citep{Laurent_2021_phemto}, GRINTA \citep{grinta} and EXPO \citep{expo}. The uncertainties shown correspond to the standard deviation from the MCMC after re-binning over each energy band (Eq.\ref{eq:rebin_polar}).
We cut the energy bands corresponding to COSI and PHEMTO in order to showcase the different polarization behavior.
As seen from the different measurable PF, PHEMTO is the most suited mission to measure the polarization transition due to the synchrotron hard-tail, while COSI could confirm the high PF expected above 500\,keV, although the source would become faint due to the synchrotron cut-off.
The XPosat and EXPO missions are on the other hand well suited to find the predicted local maximum of the PF around 10\,keV (see Fig.\ref{fig:polar_model_sync_exp_log}).
The PF prediction is also shown for the AstroSat energy range where a 15\% upper-limit was estimated, in agreement with our model \citep{Chattopadhyay_2024}. Note that the predicted polarization do not take any of the instruments characteristics into account and therefore represent pure model prediction, considered for energy ranges corresponding to the coverage of individual missions.\\  

\begin{table}[h!]
\centering
\caption{Model predictions for different energy bands with standard deviation from MCMC.}
\label{tab:models_prediction}    
{\renewcommand{\arraystretch}{1.3} 
\begin{tabular}{cc|cc}
Mission & E (keV) & PA & PF \\
\hline

XPosat & 8--30 & -19 $\pm$ 2 ° & 6.0 $\pm$ 0.7 \% \\ 
EXPO & 6--15 & -21 $\pm$ 1 ° & 5.9 $\pm$ 0.5 \% \\ 
 & 15--35 & -15 $\pm$ 4 ° & 5.7 $\pm$ 1.0 \% \\ 
COSI & 200--500 & 49 $\pm$ 4 ° & 38.3 $\pm$ 3.4 \% \\ 
 & 500--1000 & 51 $\pm$ 4 ° & 79.8 $\pm$ 0.9 \% \\ 
GRINTA & 100--200 & 41 $\pm$ 5 ° & 11.1 $\pm$ 2.1 \% \\ 
PHEMTO & 50--300 & 35 $\pm$ 7 ° & 7.6 $\pm$ 1.6 \% \\ 
 & 300--600 & 50 $\pm$ 4 ° & 61.2 $\pm$ 3.2 \% \\ 
AstroSat & 100--175 & 40 $\pm$ 5 ° & 10.2 $\pm$ 2.0 \% \\ 

\hline
\end{tabular}}
\end{table}

\section{Conclusion}

From the currently available data in the literature, we were able to set up a coherent method to fit a spectro-polarization model to the BHXB \cygx. In particular, our method can take into account the overlap of different components with different polarization, which is particularly important for the X-ray and $\gamma$-ray ranges where the energy resolution for the polarization measurements is low and total polarization can be highly energy dependent.
Using a rigorous approach of mixing polarization and simultaneous fitting of the spectrum and polarization, we are then able to separate the various components, demonstrating that such an approach is crucial. In our case, we distinguish the IC and the synchrotron hard-tail, which were historically hard to truly disentangle from spectra alone.\\

For the hard-tail, we were careful to use a full treatment of the spectro-polarization induced by a population of non-thermal electrons following a super-exponential power-law. We were able to make a proper estimation of a key parameter, the photon cut-off energy, $E_{cut}=(\Ecut^{+\Eplus}_{-\Eminus}) \times 10^{2}\ \mathrm{keV}$. This translates to a maximum electron energy of $\elecgev^{+\elecgevplus}_{-\elecgevminus}$\,GeV for typical magnetic field strength of $10^4$\,G. The Lorentz factor of the shock speed was then evaluated between $\Gamma_{sh}\approx \gammashlow - \gammashhigh$ in the Bohm regime.\\

Finally, we proposed a new explanation for the misalignment between the jet seen in radio and the polarization angle seen in $\gamma$-rays. Due to the large jet opening angle near the BH, different parts of the jets will have different viewing angle, and therefore experience different Doppler boosting and angle projections. Combined with the probable helicity of the magnetic field, this results in a shift of the polarization angle, and allows us to estimate an upper-limit on the helicity of $\helicity^{\circ}$. The sign of the helicity (left-handed for the approaching jet) also indicates that the BH spin and magnetic field have the same direction near the BH.\\

Similar studies could be performed for other transient BHXBs, in particular \emph{Swift}\,J1727$-$1613, which has simultaneous polarization measurements in soft X-rays \citep{Veledina_2023} and soft $\gamma$-rays \citep{Bouchet_2024}. This would allow us to test whether the properties of \cygx, which has a stellar wind accretion process, are similar to transient systems with Roche lobe overflow accretion.
The Crab Nebula is another promising candidate for which many polarization measurements exist \citep{Bouchet_2026}, and also contains a break in its spectrum near $\sim100$\,keV \citep{Jourdain_2019}.\\

With the recent end of operation of the INTEGRAL satellite \citep{integral_legacy}, it has become crucial to continue its legacy and probe spectro-polarization further. The COSI mission will fulfill part of this role in the near future, although it may lack the angular resolution required to distinguish sources in the Galactic center, where many BHXB reside \citep{cosi_main}.
The proposed PHEMTO and GRINTA missions could alleviate this problem and allow us to test the spectro-polarization properties of many BHXBs, with even higher precision and sensitivity \citep{Laurent_2021_phemto, grinta}.\\


\begin{acknowledgements}
TB acknowledges support from the DFG/LIS project SI 2502/6-1, project number 551127478.
JR acknowledge partial funding from the French Space Agency (CNES).
Based on observations with INTEGRAL, an ESA project with instruments and science data center funded by ESA member states (especially the PI countries: Denmark, France, Germany, Italy, Switzerland, Spain) and with the participation of Russia and the USA.
\end{acknowledgements}

%
\bibliographystyle{aa} 
\bibliography{references}

\begin{appendix}
\onecolumn

\section{Examples of log-sigmoids}

\begin{figure*}[h!]
    \centering
    \includegraphics[width=\linewidth]{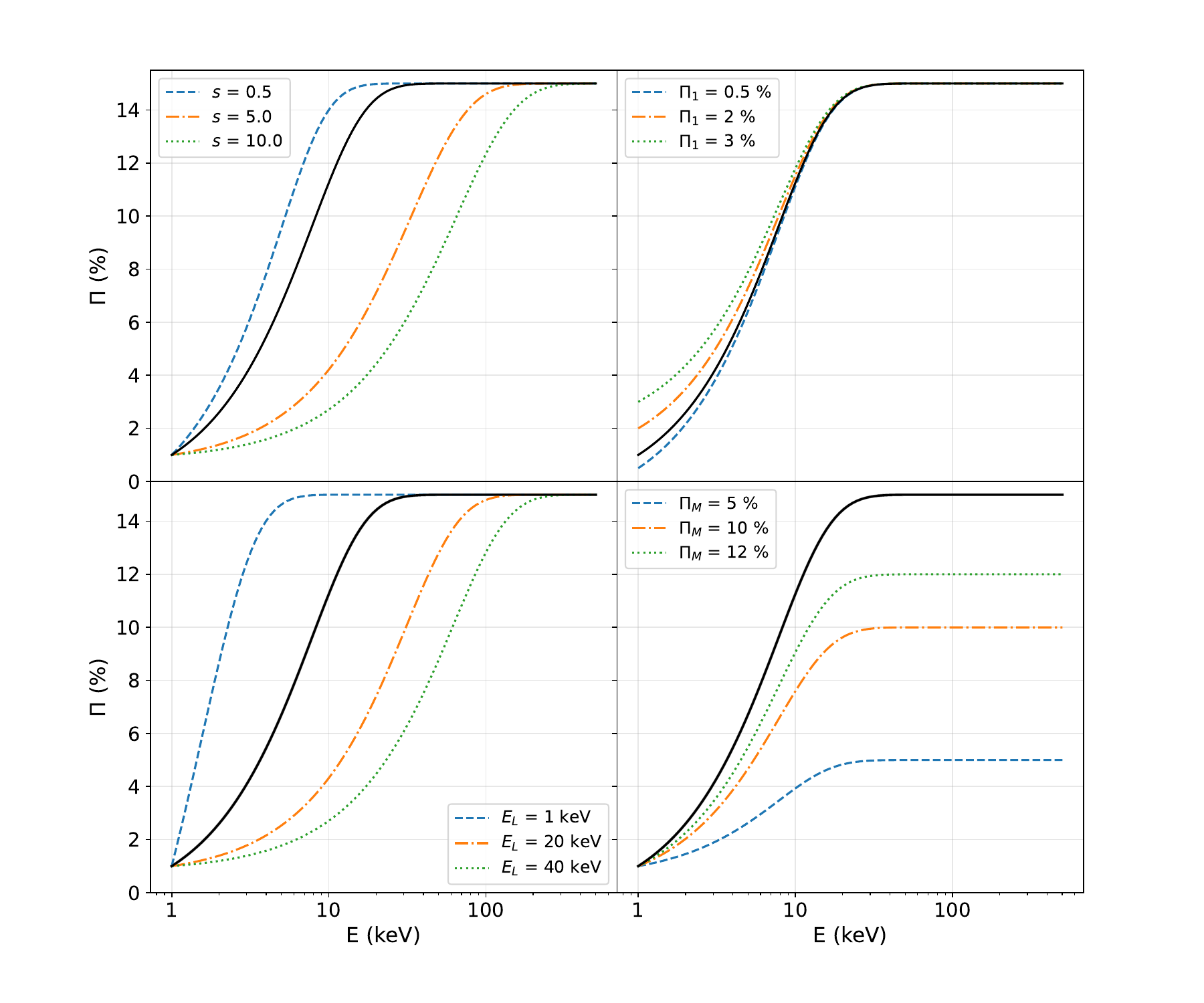}
    \caption{Examples of the log-sigmoid function for various parameters. The same default curve is shown in every plot in black with the default parameters: $\Pi_M=15\%$, $E_L=5$ keV, $\Pi_1=1\%$ and $s=1$. Each sub-plot then has one of the parameter varied while the others stay at the default value.}
    \label{fig:loglogistic_examples}
\end{figure*}

\twocolumn

\section{Numerical computation of the synchrotron polarization fraction}\label{sec:grid}

The kernel functions (Eq.\ref{eq:kernel_sync}) were approximated using the fits found by \cite{cutoff_exp_polar} in their Appendix B. The integration of the improper integrals was performed with the tanh-cosh quadrature method \citep{tanhsinh}, implemented in the \texttt{scipy} Python library. Although it is possible to simply use the local spectral index of the photon spectrum ($\Gamma(E)=-d(\ln{F})/d(\ln{E})$) to reduce the amount of computation \citep{Bjornsson_1982}, we found it deviated too significantly from the kernel integration.\\

\begin{figure}[h]
    \centering
    \includegraphics[width=\linewidth]{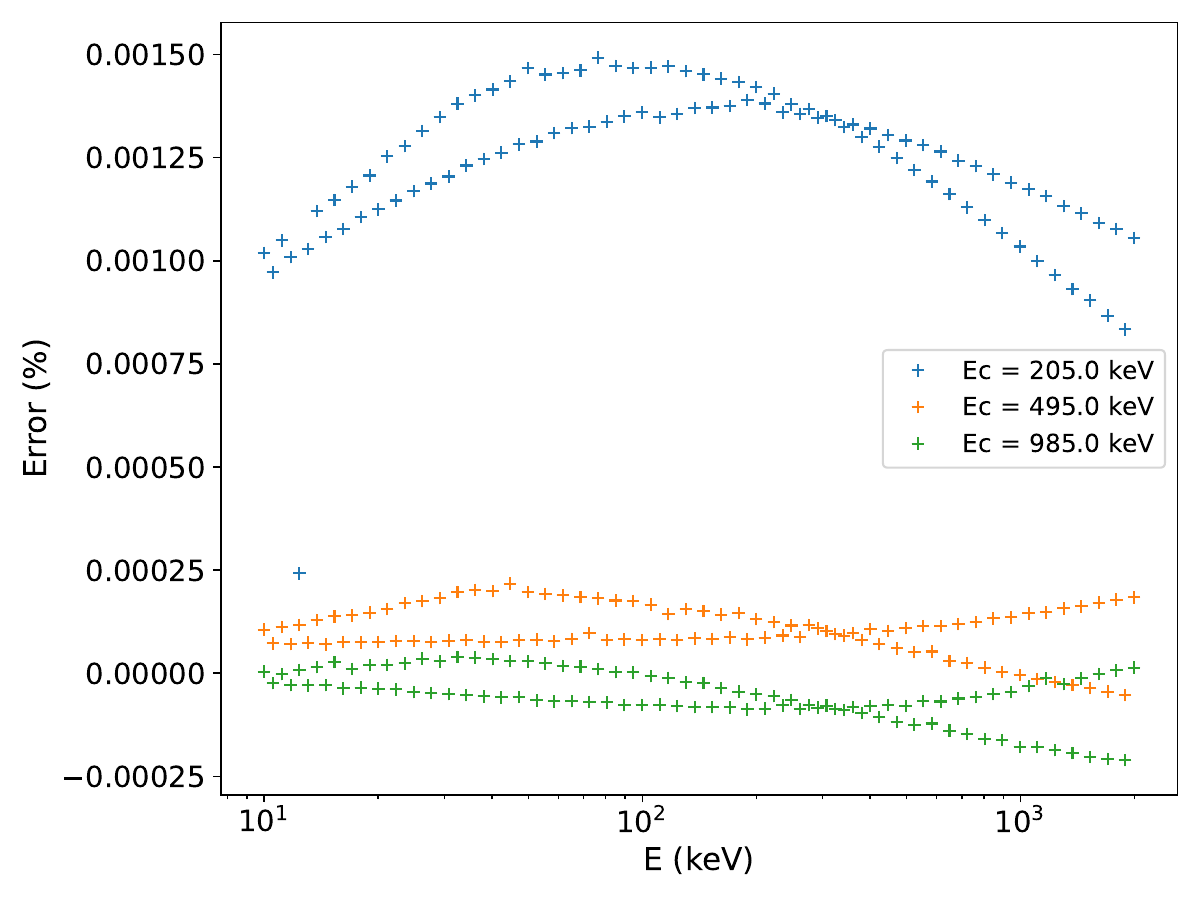}
    \caption{Difference-ratio between the grid interpolation and the complete kernel integration for different values of $E_{cut}$.}
    \label{fig:interpolation_error}
\end{figure}

The improper synchrotron integral is expensive to compute. To accelerate computation during fitting, we therefore pre-calculated the PF for 400 energies between 1 and 5000\,keV, 100 values of $E_{cut}$ between 10 and and 1000\,keV, and 60 values of $\Gamma_{sync}$ between $0.5$ and $3.5$. This generates a 3D array of the PF for relevant energies and parameters. For the fitting procedure, we first interpolate the original grid onto the energy grid used for fitting. Then, during the fitting, this new grid is interpolated between the parameters to increase the precision on $E_{cut}$ and $\Gamma_{sync}$, without having to re-interpolate the energies. Fig.\ref{fig:interpolation_error} shows the difference between the interpolation using this method and the direct integration, where we chose values of $E$ and $E_{cut}$ which are not on the original grid, and therefore truly interpolated. The error is clearly negligible, being less than $0.002\%$, while the computational time is reduced by many orders of magnitude.

\onecolumn
\section{MCMC corner plot}\label{sec:mcmc}

\begin{figure*}[h!]
    \centering
    \includegraphics[width=\linewidth]{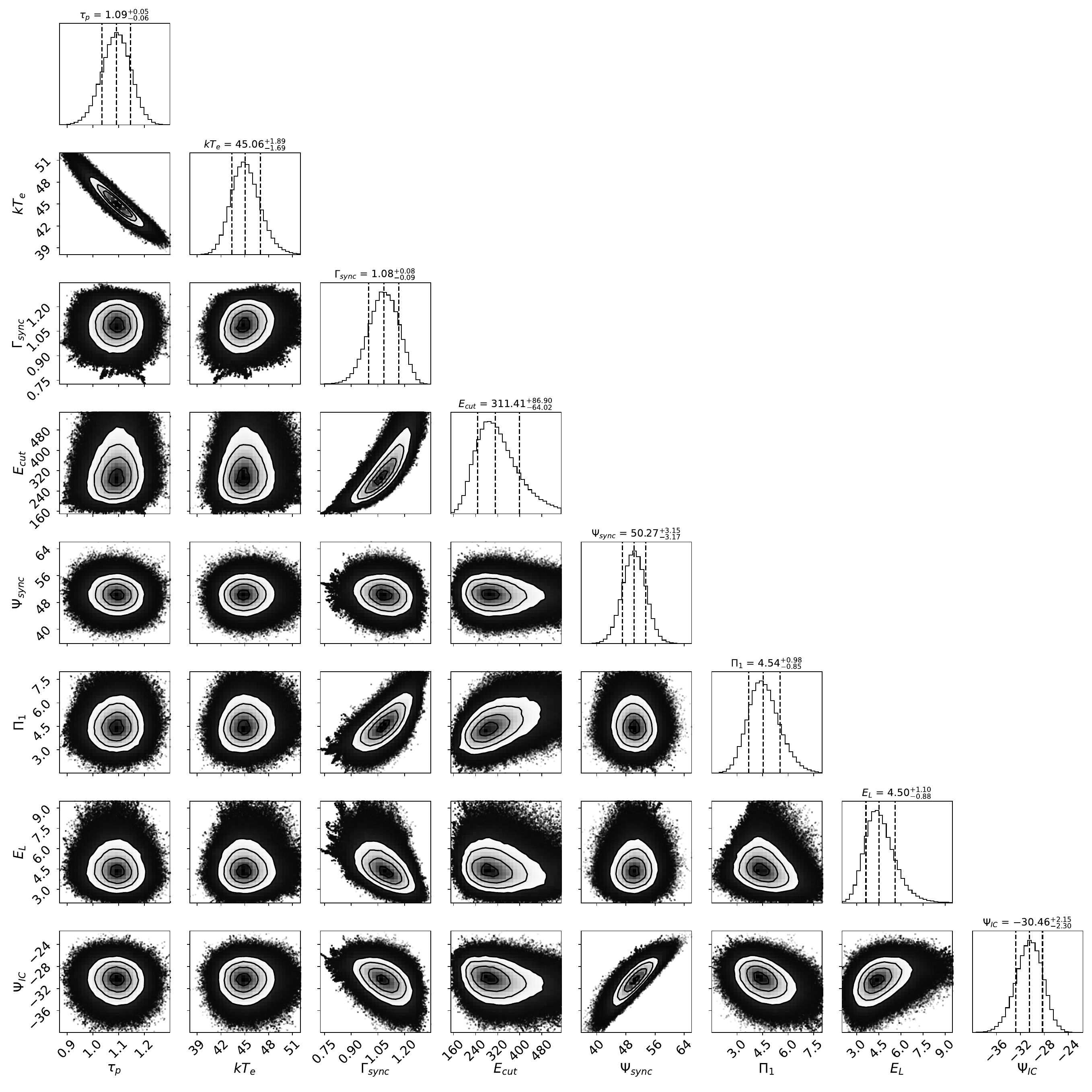}
    \caption{Corner plot of the MCMC showing the 2D correlation between the main parameters of interest. At the top of each column, the 1D distribution is also shown, with the $50 \pm 34$\% quantiles.} 
    \label{fig:emcee_corner_best}
\end{figure*}

\twocolumn

\end{appendix}
\end{document}